\documentclass[onecolumn]{aa} 
\usepackage{amsmath} 
\usepackage[pdftex]{graphicx} 
\usepackage{txfonts}
\usepackage{url}
\begin{document}
\title{``TNOs are Cool": A Survey of the Transneptunian Region IV }

\subtitle{Size/albedo characterization of 15 scattered disk and detached objects observed with \textit{Herschel} Space Observatory-PACS\thanks{\textit{Herschel} is an ESA space observatory with science instruments provided by European--led Principal Investigator consortia and with important participation from NASA. PACS: The Photodetector Array Camera and Spectrometer is one of \textit{Herschel's} instruments.}}

   \author{P. Santos-Sanz\inst{1}
		   \and	
		   E. Lellouch\inst{1}
		   \and
		   S. Fornasier\inst{1,2}
		   \and
		   C. Kiss\inst{3}
		   \and
		   A. Pal\inst{3}
		   \and
		   T.G. M\"uller\inst{4}
		   \and
		   E. Vilenius\inst{4}
		   \and		   
		   J. Stansberry\inst{5}
		   \and
		   M. Mommert\inst{6}
		   \and 
		   A. Delsanti\inst{1,7}
		   \and	
		   M. Mueller\inst{8,9}
		   \and	
		   N. Peixinho\inst{10,11}
		   \and
		   F. Henry\inst{1}
		   \and		   
		   J.L. Ortiz\inst{12}
		   \and
		   A. Thirouin\inst{12}
		   \and
		   S. Protopapa\inst{13}        
		   \and
		   R. Duffard\inst{12}
		   \and		   
		   N. Szalai\inst{3}
		   \and		   	
		   T. Lim\inst{14}
		   \and
		   C. Ejeta\inst{15}		   
		   \and		   
		   P. Hartogh\inst{15}	   
		   \and		   
  		   A.W. Harris\inst{6}
		   \and
		   M. Rengel\inst{15}}

\offprints{P. Santos-Sanz: pablo.santos@obspm.fr}

\institute{LESIA-Observatoire de Paris, CNRS, UPMC Univ. Paris 6, Univ. Paris-Diderot, 
5 Place J. Janssen, 92195 Meudon Cedex, France.
 \email{pablo.santos@obspm.fr} 
 \and
 Univ. Paris Diderot, Sorbonne Paris Cit\'{e}, 4 rue Elsa Morante, 75205 
Paris, France.
 \and 
 Konkoly Observatory of the Hungarian Academy of Sciences, Budapest, Hungary.
\and
 Max--Planck--Institut f\"ur extraterrestrische Physik (MPE), Garching, Germany.
 \and
 University of Arizona, Tucson, USA.
\and
 Deutsches Zentrum f\"ur Luft- und Raumfahrt (DLR), Institute of Planetary Research, Berlin,
Germany.
\and
 Laboratoire d'Astrophysique de Marseille, CNRS \& Universit\'e de Provence, Marseille.
\and
SRON LEA / HIFI ICC, Postbus 800, 9700AV Groningen, Netherlands.
\and
UNS-CNRS-Observatoire de la C\^ote d${'}$Azur, Laboratoire Cassiop\'ee, BP 4229, 06304 Nice Cedex 04, France.
\and
Center for Geophysics of the University of Coimbra, Av. Dr. Dias da Silva, 3000-134 Coimbra, Portugal
\and 
Astronomical Observatory, University of Coimbra, Almas de Freire, 3040-004 Coimbra, Portugal
\and
 Instituto de Astrof\'{\i}sica de Andaluc\'{\i}a (CSIC), Granada, Spain.
\and
University of Maryland, USA.
\and
Space Science and Technology Department, Rutherford Appleton
Laboratory, Chilton, Didcot, Oxon UK, OX11 0QX
\and
 Max-Planck-Institut f\"ur Sonnensystemforschung (MPS), Katlenburg-Lindau, Germany.
}
   
   \date{Received 29 November 2011 / Accepted 24 January 2012}

 
  \abstract
   {Physical characterization of Trans-Neptunian objects, a primitive population of the outer solar system,  may provide constraints on their formation and 
evolution.}
   {The goal of this work is to characterize a set of 15 scattered disk (SDOs) and detached objects, in terms of their size, albedo,
and thermal properties.}
   {Thermal flux measurements obtained with the \textit{Herschel}-PACS instrument at 70, 100 and 160 $\mu$m, and whenever applicable, with \textit{Spitzer}-MIPS
at 24 and 70 $\mu$m, are modeled with radiometric techniques, in order to derive the objects' individual size, albedo and when possible
beaming factor. Error bars are obtained from a Monte-Carlo approach. We look for correlations between these and other
physical and orbital parameters.}
   {Diameters obtained for our sample range from 100 to 2400 km, and the geometric albedos (in V band) vary from 3.8 \% to 84.5 \%. The unweighted mean V geometric albedo for the whole sample is 11.2 \% (excluding Eris); 6.9 \% for the SDOs, and 17.0$~\%$ for the detached objects (excluding Eris). We obtain new bulk densities for three binary systems: Ceto/Phorcys, Typhon/Echidna and Eris/Dysnomia. Apart from correlations clearly due to observational bias, we find significant correlations between albedo and diameter (more reflective objects being bigger), and between albedo, diameter and perihelion distance (brighter and bigger objects having larger perihelia). We discuss possible explanations for these correlations.}
   {}

   
\titlerunning{``TNOs are Cool": A Survey of the Transneptunian Region. IV.}

   \maketitle

%

\section{Introduction}

Since the detection of the first Trans-Neptunian Object (TNO) 20 years ago --1992 QB$_{1}$-- (Jewitt and Luu 1992) besides Pluto, 
about 1550 TNOs/Centaurs have been discovered (as of October 2011). Different dynamical classes, according to their orbital parameters, 
have been identified within this distant icy population: classical objects with moderate eccentricity orbits; resonant objects trapped 
in mean motion resonances with Neptune (Plutinos, in the 3:2 resonance with Neptune, represent the most populated
resonant group, \cite{2000prpl.conf.1201J}); scattered disk objects (SDOs) with high eccentricities and sometimes high inclinations, 
resulting from close encounters with Neptune; detached objects, whose perihelia are well outside Neptune's orbit 
(\cite{2002Icar..157..269G}), and which likely constitute a large population in the outer Kuiper Belt;
centaurs, which reside between the orbits of Jupiter and Neptune, represent TNOs scattered towards the inner solar system,
and so are usually considered as Kuiper Belt Objects (KBOs). Although it is convenient to classify objects by ``populations",
the limits between some of these populations (i.e. between SDOs, centaurs, high inclination/high eccentricity classicals, detached, etc...) 
are often ``blurred" and may ultimately rely on definition (\cite{2008ssbn.book...43G}). The present paper deals with 15 SDO or detached objects, for which we first 
specify the adopted definition.

The terminology ``scattered disk'' was originally used for TNOs with orbits characterized by both a high eccentricity and a perihelion 
close to Neptune. It is believed that the majority of SDOs have unstable orbits due to close encounters with Neptune, 
but some of them could have survived since the origin of the solar system (\cite{1997Sci...276.1670D}). Alternatively, 
the origin of the scattered disk could be related to other effects, including sweeping resonances that move the objects to this region (\cite{1995AJ....110..420M}), 
passing star (or stars) (\cite{2004AJ....128.2564M} and references therein). Several approaches to the classification of objects as SDOs numerically integrate their orbits backward in time over dozens of millions of years (\cite{2005AJ....129.1117E}, \cite{2008ssbn.book...43G}). Here we adopt the Gladman et al. (2008) classification, which defines SDOs as those objects, independently of their origin, 
that are currently scattering actively off Neptune. SDOs are characterized by the rapid variation of semi-major axis in the orbit integrations. Specifically, Gladman at al. (2008) adopt a criterion similar to Morbidelli et al. 
(2004): the object is classified as an SDO if its semimajor-axis presents a variation $\geq 1.5$ AU during a 10-m.y. backward integration. The exact value used (1-2 AU variation) makes little difference, as SDOs undergo large semimajor-axis changes in short times. 
Then, SDOs in this nomenclature might be called ``scattering" objects instead of ``scattered'' objects. 
Following this classification SDOs exist over a large semi-major axis range, not only restricted to $a >50$ AU; rather the SDO population extends down to the Centaur region (beginning at $a$~=~30 AU). At the other edge the inner Oort cloud would begin at very large semimajor-axes ($a > 2000$ AU, \cite{2008ssbn.book...43G}).

\emph{Detached} objects (\cite{2006ssu..book..267D}) are also known as \emph{``extended scattered disk objects"} (ESDOs),
though it is preferable to use the term ``detached" because not all these objects have necessarily been placed in this region 
through scattering. Detached objects have orbits with perihelion distances unaffected (``decoupled") from the giant planets, in particular Neptune. 
This population may include inner Oort cloud bodies or transitional objects between SDOs and inner Oort cloud objects. The limits in parameter space are unknown, and are expanding with the new discoveries of TNOs at larger and larger semimajor axes --e.g. a= 965 AU for 2006 SQ$_{372}$-- (\cite{2002Icar..157..269G}). Elliot et al. (2005) adopted the Tisserand parameter (T) relative to Neptune to 
define the detached population. A problem with this definition is the large number of high inclination TNOs within the classical population. In the Gladman et al. (2008) scheme, numerical integrations separate SDOs from detached TNOs. A remaining issue 
is to decide where the detached population ends towards low eccentricities; Gladman et al. (2008) adopt a lower bound of \textit{e}~=~0.24. 
Possible processes that placed the detached objects on these orbits are currently being investigated (e.g. \cite{Gomes2008}, \cite{Kenyon2008}, 
\cite{Duncan2008}, etc).

Determining KBO sizes and albedos is not a straightforward task. Measurements from direct imaging are available
for only a handful of them (\cite{2004AJ....127.2413B}, \cite{Brown2006}). Most recently, sizes have become
available from stellar occultations (\cite{2010Natur.465..897E}, \cite{2011Natur.478..493S}). Obtaining
size/albedo values for a larger sample is best achieved from thermal radiometry, whereby thermal and optical
observations are combined within a thermal model. After the pioneering measurements on two objects from ISO (\cite{2000ApJ...534..446T}) and
ground-based observations on a few others (see \cite{stansberry08} and references therein), 
{\em Spitzer} provided the first large such dataset, with about 60
Centaurs/KBOs measured at 24 and/or 70 $\mu$m (\cite{2005ApJ...624L..53C}, Grundy et al. 2005 \& 2007, Stansberry et al. 2006 \& 2008, \cite{Brucker2009}). This approach (multi-wavelength photometry) also constrains the thermal regime of the object, phenomenologically
described by the ``beaming factor" (\cite{1974AJ.....79..892J}, \cite{1989aste.conf..128L}, \cite{Harris98}). The launch of \textit{Herschel} Space Observatory in 2009 offered a new tool for this purpose, providing photometry over 6 bands from 70-500
$\mu$m. Early results on a handful of objects were presented in \cite{2010A&A...518L.146M}, \cite{Lim2010}, and \cite{2010A&A...518L.147L}. Beyond their intrinsic scientific value, albedos
are needed to properly interpret reflectance spectra. Furthermore, correlations between size, albedo, color, composition and
orbital parameters may provide diagnostics on the dynamical, collisional
and physical history of KBOs. The size distribution of TNOs is also a
diagnostic relating to that history, but in the absence of measured sizes
the distribution has typically been estimated from the luminosity function
and an assumed albedo, an uncertain step given that the average albedos seem to be different
for the different dynamical populations (\cite{2005Icar..176..184G}, \cite{Brucker2009}). Instead, and to constrain Kuiper belt formation/evolution models  it is desirable to directly determine the size 
distribution of the different populations, and infer their individual total masses.
Large bodies retain information about the angular momentum of the presolar nebula and 
their size distribution is dominated by accretion. On the other hand, small bodies --in the 50 to 100 km range-- are the result 
of collisional evolution (\cite{2008ssbn.book...71P}). 

A large program (370 h of telescope time), named ``TNOs are Cool: A Survey of the Transneptunian Region" (\cite{2009EM&P..105..209M}), has been awarded on \textit{Herschel} 
to measure the thermal flux of about 140 KBOs/Centaurs. We present
here results on 15 SDO and detached objects. These observations lead to new sizes/albedos for 9 objects and improved
results on 6 others. Results on 19 classical and 18 Plutinos are presented in companion
papers by Vilenius et al. (2012) and Mommert et al. (2012), respectively.  


\section{Observations}
\label{observations}

The \textit{Herschel} Space Observatory (\cite{2010A&A...518L...1P}) is a 3.5 m telescope launched on 14 May 2009, covering
the far-infrared range by means of three focal plane instruments. The ``TNOs are Cool" program consists of photometry of $\sim$140
objects, mostly with the Photodetector Array Camera and Spectrometer (PACS, \cite{2010A&A...518L...2P}).

The PACS instrument has 3 photometric bands centered at 70, 100 and 160 $\mu m$. These bands (termed \textit{`blue'}, \textit{`green'} and \textit{`red'}, respectively) 
cover the 60-85 $\mu m$,  85-130 $\mu m$, 130-210 $\mu m$ ranges. With two bolometer arrays, PACS can observe a field in two bands at a time: blue/red or green/red. The red detector is a $32 \times 16$ pixel array, while the blue/green is $64 \times 32$ pixels. The 3 bands image the same field of view of $3.50' \times 1.75'$. Projected onto the sky, pixel sizes are $6.4\arcsec \times 6.4\arcsec$ (red band) and $3.2\arcsec \times 3.2\arcsec$ (blue/green bands), respectively.

The observations discussed here were performed using the scan map mode. This mode was selected due to its better overall performance compared to the point-source (chop-nod) mode (\cite{2010A&A...518L.146M}). Scan maps are performed by slewing the spacecraft at a constant speed along parallel lines, or ``legs". 
Our observations were typically performed at medium speed ($20\arcsec/s$). We use 10 scan legs in each map, separated by $4\arcsec$. The length of each leg is $3.0\arcmin$. Each such scan-map is repeated two to five times, depending on the expected brightness of the objects (see below). Within a single scan map, either the blue-red or the green-red channel combination is selected. For any band selection, two individual maps are made using two scanning directions with respect to the detector array: $70\degr$ and $110\degr$, allowing to fit the scan pattern on the rectangular array in an optimum way.

The ``TNOs are cool'' target list was constructed by predicting the
thermal fluxes in the PACS bands for the known TNOs using the ``hybrid
standard thermal model'' (STM -- see Sect. \ref{thermalmodel}). The flux predictions were
based on absolute visual magnitudes (H) from the Minor Planet Center (MPC),
an assumed visual geometric albedo of 8\% (based on previous studies using
{\em Spitzer}), and geometric circumstances appropriate to the expected
observing epoch (c. 2010). For objects with known albedos (for this
study those were Eris: 70\%, Typhon: 5\%, 1996~TL$_{66}$: 4\%, and
2002~PN$_{34}$: 5\%) we used those values rather than the 8\%
assumption. Objects with the largest predicted thermal fluxes were
selected as potential targets; that list was further refined by rejecting
objects with predicted positional uncertainties $< 5\arcsec$. The
objects presented in this paper are a subset of the targets selected
in the above manner. Once the targets were selected, integration times
were chosen to provide signal-to-noise ratio (SNR) values of $\sim 10$
in the blue and green channels. Integration times for individual
observations were tailored by varying the repetition factor for the scan
maps described earlier. Note that the values of H from the MPC were used
for planning purposes only, and can differ significantly (by up to 0.5 mag)
from the values ultimately used in our analysis. 

It is important to note that the target selection process introduces a bias towards brighter (low H values) objects. A consequence
of the assumed and generally fixed albedo is that the object sample favors large objects at large heliocentric distances ($r_h$). In addition, the object
sample is affected by a discovery bias that: (i) favors high-albedo objects at high $r_h$, and (ii) introduces
a correlation between the perihelion ($q$) and the heliocentric distance, because objects 
are most easily discovered near perihelion and haven't moved much since discovery. These biases and correlations
must be kept in mind when studying correlations between size, albedo and other physical and
orbital parameters (see Sect. \ref{results_and_discussion}). As for the present SDOs/Detached sample, objects are listed in Table \ref{tableOBS}, including relevant orbital parameters, absolute magnitudes in V-band, rotational periods and amplitudes, taxonomic color class (when known), existence or not of satellites, and dynamical classification following Gladman et al. (2008). Observational circumstances are described in more details in Table \ref{tableOBSIDs}.

Each target was visited twice, with identical observations repeated in both visits. During the first visit, we grouped 4 maps to be observed in sequence, corresponding to the two scan directions and the two channel combinations. The observation geometry is such that on the final combined map, the central $\sim 50\arcsec \times 50\arcsec$ region has high coverage. Photometric accuracy is usually limited by instrumental noise  in the blue channel
and by background confusion noise in the red, while either of the two sources of noise may dominate in the green\footnote{PACS AOT Release Note: PACS Photometer Point/Compact Source Mode, 2010, PICC-ME-TN-036, Version 2.0, custodian T. M{\"u}ller,
\url{http://herschel.esac.esa.int/twiki/bin/view/Public/PacsCalibrationWeb}}. The timing of the observations, i.e.~the selection of the visibility window, was optimized to ensure the lowest far-infrared confusion noise circumstances~(\cite{Kiss2005}), in such a way that the estimated SNR due to confusion noise has its maximum in the green channel. The timing of the second visit was calculated such that the target moved 30-$50\arcsec$ between the two visits. This ensures that the target position during the second visit is still within the high-coverage area of the map from the first visit. At the same time, this enables us to determine and subtract the background in the individual maps, so as to obtain a clean combined image where we extract the final flux (see Sect. \ref{DataRed}). 

\scriptsize
\begin{table*}

\caption{Orbital parameters, absolute magnitudes, B-R colors, photometric variation, taxonomy, dynamical classification and multiplicity
of the observed objects} 

\label{tableOBS} 

\centering 
\scalebox{0.9}{%
\begin{tabular}{l c c c c l c c c c c} 

\hline\hline 

Object	&	a 	&	q 	&	i 	&	e	&	H$_{V}$	&	B-R & P	&	$\Delta$m$_{R}$	&	Taxon.	&	Class.	\\
	&	[AU]	&	[AU]	&	[deg]	&		&[mag]& [mag]&	[h]	&	[mag]	&		&		\\


\hline 

1996 TL$_{66}$	&	84.5	&	35.0	&	23.9	&	0.59	&	5.39$\pm$0.12$^{a,b,c}$ & 1.11$\pm$0.07$^{c,q,r,s}$	&	12.0$^{x}$	&	$<$0.12, $<$0.06$^{x,y}$	&	BB	&	SDO	\\
2001 FP$_{185}$	&	212.0	&	34.3	&	30.8	&	0.84	&	6.39$\pm$0.07$^{a,b}$ & 1.40$\pm$0.06$^{b,t}$	&	...	&	$<$0.06$^{z}$	&	IR	&	SDO	\\
2002 PN$_{34}$	&	31.2	&	13.4	&	16.6	&	0.57	&	8.66$\pm$0.03$^{a,d}$ & 1.28$\pm$0.02$^{t}$	&	8.45/10.22$^{\alpha}$	&	0.18$\pm$0.04$^{\alpha}$	&	BB-BR	&	SDO	\\
2002 XU$_{93}$	&	66.5	&	21.0	&	77.9	&	0.69	&	8.11$\pm$0.10$^{e,f}$ & 1.20$\pm$0.02$^{e}$	&	...	&	...	&	...	&	SDO	\\
2007 OR$_{10}$	&	67.1	&	33.6	&	30.7	&	0.50	&	\textbf{1.96$\pm$0.16$^{g}$}	& ... &	...	&	...	&	...	&	SDO	\\
2007 RW$_{10}$	&	30.4	&	21.2	&	36.0	&	0.30	&	\textbf{6.39$\pm$0.61$^{g}$}	& ... & ...	&	...	&	...	&	SDO	\\
(2003 FX$_{128}$) Ceto*$^{\delta}$	&	99.7	&	17.8	&	22.3	&	0.82	&	6.54$\pm$0.06$^{h,i}$ & 1.42$\pm$0.04$^{t}$ &	4.43$^{n}$	&	0.13$\pm$0.02$^{n}$	&	...	&	SDO	\\
(2002 CR$_{46}$) Typhon*$^{j}$	&	37.6	&	17.5	&	2.4	&	0.54	&	7.72$\pm$0.04$^{a,d,f,h,j,k,m}$ & 1.29$\pm$0.07$^{m,t}$ &	9.67/19.34$^{y,\alpha}$	&	0.06$\pm$0.01$^{y,\alpha}$	&	BR	&	SDO	\\
1999 KR$_{16}$	&	48.7	&	33.9	&	24.9	&	0.30	&	\textbf{5.37$\pm$0.08$^{n,o}$} & 1.87$\pm$0.07$^{c,u,v}$ &	5.93/11.86$^{n}$	&	0.18$\pm$0.04$^{n}$	&	RR	&	DO	\\
2003 FY$_{128}$	&	49.2	&	37.0	&	11.8	&	0.25	&	5.09$\pm$0.09$^{k}$ & 1.65$\pm$0.02$^{e}$ &	8.54$^{y}$	&	0.15$\pm$0.01$^{y}$	&	BR 	&	DO	\\
2005 QU$_{182}$	&	114.0	&	37.0	&	14.0	&	0.68	&	3.80$\pm$0.32$^{f}$ & ... & ...	&	...	&	...	&	DO	\\
2005 TB$_{190}$	&	76.5	&	46.2	&	26.4	&	0.40	&	4.40$\pm$0.11$^{f}$ & 1.54$\pm$0.03$^{e}$ &	...	&	...	&	...	&	DO	\\
2007 OC$_{10}$	&	50.0	&	35.5	&	21.7	&	0.29	&	5.43$\pm$0.10$^{f,i}$ & ... & ... &	...	& ...	&	DO	\\
2007 UK$_{126}$*$^{\zeta}$	&	74.0	&	37.7	&	23.4	&	0.49	&	3.69$\pm$0.10$^{k}$ & ... & ...	&	...	&	...	&	DO	\\
(2003 UB$_{313}$) Eris*$^{\epsilon}$	&	68.1	&	38.6	&	43.8	&	0.43	&	-1.12$\pm$0.03$^{d,p}$	& 1.21$\pm$0.09$^{d,w}$ &	13.69/28.08$^{\beta}$	&	$<$0.1, 0.01$^{\beta,\gamma}$	&	BB	&	DO	\\
	&		&		&		&		&			&	& 32.13/25.92$^{\beta,\gamma}$	&		&		&		\\

\hline 

\end{tabular}}

\begin{flushleft}
\footnotesize{ \textbf{*} Indicates that the object is a known binary/multiple system. \textbf{H$_{V}$ [mag]:} average visual magnitude obtained from papers referenced below. Numbers in bold indicate $H_{R}$ magnitudes instead of $H_{V}$. See text for details about the absolute magnitude estimation. \textbf{B-R [mag]} colors (when available). \textbf{P [h]} single or double-peaked rotational period. \textbf{$\Delta$m$_{R}$ [mag]:} lightcurve amplitude. \textbf{Taxon.}: taxonomic color class (Perna et al. 2010 and references therein). \textbf{Class.}: dynamical classification following  Gladman et al. 2008. SDO = scattered disc object, DO = detached object. \textbf{References:} $^{a)}$ \cite{2005Icar..179..523R}; $^{b)}$ \cite{2005Icar..174...90D}; $^{c)}$ \cite{2001AJ....122.2099J}; $^{d)}$ \cite{2007AJ....133...26R}; $^{e)}$ \cite{2010AJ....139.1394S}; $^{f)}$ Derived from MPC V-data; $^{g)}$ Derived from MPC R-data; $^{h)}$ \cite{2009Icar..200..292B}; $^{i)}$ Perna \& Dotto 2011, priv. comm.; $^{j)}$ \cite{2008Icar..197..260G}; $^{k)}$ \cite{2010A&A...510A..53P}; $^{m)}$ \cite{2004Icar..170..153P}; $^{n)}$ \cite{2002AJ....124.1757S}; $^{o)}$ \cite{2003EM&P...92..201B}; $^{p)}$ \cite{2007AJ....134..787S}; $^{q)}$ \cite{1998Natur.392...49T}; $^{r)}$ \cite{1998AJ....115.1667J}; $^{s)}$ \cite{1999Icar..142..476B}; $^{t)}$ \cite{2003ApJ...599L..49T}; $^{u)}$ \cite{2002ApJ...566L.125T}; $^{v)}$ \cite{2002A&A...395..297B}; $^{w)}$ Tegler 2011, priv. comm.; $^{x)}$ \cite{2006A&A...447.1131O}; $^{y)}$ \cite{2010A&A...522A..93T}; $^{z)}$ \cite{2003EM&P...92..207S}; $^{\alpha)}$ \cite{2003A&A...407.1149O}; $^{\beta)}$ \cite{2008A&A...479..877D}; $^{\gamma)}$ \cite{2008Icar..198..459R}; $^{\delta)}$ \cite{2007Icar..191..286G}; $^{\epsilon)}$ \cite{2007Sci...316.1585B}.}; $^{\zeta)}$ \cite{2009DPS....41.4707N}
\end{flushleft}

\end{table*}

\normalsize

\begin{table*}

\caption{Individual observational circumstances} 

\label{tableOBSIDs} 
 
\centering 

\begin{tabular}{r c c c c c c} 

\hline\hline 

Object	&	OBSIDs	&	Dur. [min]	&	Mid-time	&	$r_h$ [AU]	&	$\Delta$[AU]	&	$\alpha$[deg]	\\


\hline 

(15874) 1996 TL$_{66}$	&	1342190953-0956	&	39.7	&	22-Feb-2010 10:57:34	&	35.7245	&	35.8103	&	1.59	\\
(15874) 1996 TL$_{66}$	&	1342191029-1032 &	39.7	&	23-Feb-2010 12:09:47	&	35.7250	&	35.8289	&	1.59	\\
(82158) 2001 FP$_{185}$	&	1342211422-1425	&	97.4	&	23-Dec-2010 08:41:10	&	34.9353	&	35.2646	&	1.53	\\
(82158) 2001 FP$_{185}$	&	1342211528-1531	&	97.4	&	23-Dec-2009 21:36:49	&	34.9356	&	35.2562	&	1.53	\\
(73480) 2002 PN$_{34}$	&	1342213067-3070	&	41.0	&	18-Jan-2011 14:18:48	&	17.5092	&	17.8769	&	2.99	\\
(73480) 2002 PN$_{34}$	&	1342213089-3092	&	41.0	&	18-Jan-2011 21:31:02	&	17.5098	&	17.8821	&	2.98	\\
(127546) 2002 XU$_{93}$	&	1342204211-4214	&	59.8	&	09-Sep-2010 14:56:03	&	21.1063	&	21.4432	&	2.58	\\
(127546) 2002 XU$_{93}$	&	1342204240-4243	&	59.8	&	09-Sep-2010 21:19:45	&	21.1064	&	21.4399	&	2.58	\\
2007 OR$_{10}$	&	1342220081-0084	&	78.6	&	07-May-2011 02:16:22	&	86.3285	&	86.5992	&	0.65	\\
2007 OR$_{10}$	&	1342220272-0275	&	78.6	&	09-May-2011 23:59:21	&	86.3305	&	86.5534	&	0.66	\\
2007 RW$_{10}$	&	1342213219-3222	&	59.8	&	24-Jan-2010 05:43:55	&	27.4552	&	27.8385	&	1.90	\\
2007 RW$_{10}$	&	1342213270-3273	&	59.8	&	24-Jan-2011 21:05:07	&	27.4558	&	27.8495	&	1.89	\\
(65489) Ceto	&	1342202877-2880	&	59.8	&	11-Aug-2010 20:15:24	&	31.6508	&	31.8552	&	1.81	\\
(65489) Ceto	&	1342202910-2913	&	59.8	&	12-Aug-2010 11:00:59	&	31.6524	&	31.8666	&	1.80	\\
(42355) Typhon	&	1342210596-0599	&	59.8	&	30-Nov-2010 21:13:47	&	18.2160	&	18.5731	&	2.90	\\
(42355) Typhon	&	1342210624-0627	&	59.8	&	01-Dec-2010 03:50:07	&	18.2162	&	18.5690	&	2.91	\\
(40314) 1999 KR$_{16}$	&	1342212814-2817	&	97.4	&	17-Jan-2011 20:35:59	&	35.7593	&	36.0633	&	1.51	\\
(40314) 1999 KR$_{16}$	&	1342213071-3074	&	97.4	&	18-Jan-2011 15:51:40	&	35.7589	&	36.0497	&	1.51	\\
(120132) 2003 FY$_{128}$	&	1342212770-2773	&	59.8	&	16-Jan-2011 19:01:01	&	38.8406	&	38.6763	&	1.45	\\
(120132) 2003 FY$_{128}$	&	1342213107-3110	&	59.8	&	19-Jan-2011 05:03:50	&	38.8416	&	38.6356	&	1.44	\\
2005 QU$_{182}$	&	1342212619-2622	&	41.0	&	14-Jan-2011 14:46:12	&	48.8994	&	49.1349	&	1.13	\\
2005 QU$_{182}$	&	1342212696-2699	&	41.0	&	15-Jan-2011 14:01:49	&	48.9008	&	49.1522	&	1.12	\\
(145480) 2005 TB$_{190}$	&	1342221729-1732	&	78.6	&	27-May-2011 03:17:15	&	46.3134	&	46.5135	&	1.24	\\
(145480) 2005 TB$_{190}$	&	1342221782-1785	&	78.6	&	28-May-2011 13:15:54	&	46.3132	&	46.4900	&	1.25	\\
2007 OC$_{10}$	&	1342206671-6674	&	59.8	&	16-Oct-2010 21:52:04	&	35.5187	&	35.1842	&	1.54	\\
2007 OC$_{10}$	&	1342206698-6701	&	59.8	&	18-Oct-2010 11:48:52	&	35.5189	&	35.2090	&	1.55	\\
(229762) 2007 UK$_{126}$	&	1342202277-2280	&	41.0	&	08-Aug-2010 18:09:36	&	44.9448	&	45.1702	&	1.27	\\
(229762) 2007 UK$_{126}$	&	1342202324-2327	&	41.0	&	09-Aug-2010 13:05:14	&	44.9440	&	45.1572	&	1.27	\\
(136199) Eris	&	1342199487-9490	&	59.8	&	30-Jun-2010 21:01:08	&	96.6522	&	96.8807	&	0.59	\\
(136199) Eris	&	1342199753-9756	&	59.8	&	03-Jul-2010 19:45:58	&	96.6518	&	96.8326	&	0.60	\\

\hline 

\end{tabular}

\footnotesize{
\textbf{OBSIDs:} \textit{Herschel} internal observation IDs. \textbf{Dur. [min]:} total duration of the 4 OBSIDs observations for the red band. For the blue and green channels the time is half of this value. \textbf{Mid-time:} Mean date and UT time of the observation. \textbf{$r_h$ [AU]:} heliocentric distance (AU) at mid-time. \textbf{$\Delta$ [AU]:} distance object-\textit{Herschel} (AU) at mid-time. \textbf{$\alpha$[deg]:} phase angle -degrees.}

\end{table*}


\section{Data reduction and photometry}\label{dataphotom}

In this section, we first describe the reduction process applied to the PACS data in order to obtain what we call ``single" and ``combined" maps, and then the photometric techniques used to determine the target fluxes and associated uncertainties in each band. Further details of the data reduction are given in Appendix A. 


\subsection{Data reduction}
\label{DataRed}

The data reduction of the PACS maps was performed within the \textit{Herschel} Interactive Processing Environment (HIPE\footnote{HIPE is a joint development by the \textit{Herschel} Science Ground Segment Consortium, consisting of ESA, the NASA \textit{Herschel Science Center}, and the HIFI, PACS and SPIRE consortia members, see: http://herschel.esac.esa.int/DpHipeContributors.shtml}, version 6.0.2044) by means of adapted standard HIPE scripts. Since our targets move slowly (a few $\arcsec$/hour), and the total observation time per scanning direction in one visit is only 10-25 min, we do not correct for the apparent motion.

After the application of the HIPE scripts we obtain one ``single" map per visit, filter, and scan direction (i.e. in total 8 maps per object in the red, and 4 maps in the blue/green). The final re-sampled pixel scale of the single maps are 1.1$\arcsec$/pixel, 1.4$\arcsec$/pixel, and 2.1$\arcsec$/pixel for the blue (70 $\mu m$), green (100 $\mu m$) and red (160 $\mu m$) channels respectively.

We use these single maps to generate final ``combined" maps on which the photometry will be performed. To generate the combined 
maps, we combine the two visits to the target to determine the background map and subtract it from each single map. Finally we co-add all the background-subtracted maps in the co-moving frame of the target. Similar methods were used to subtract background sources in \textit{Spitzer} TNO observations, and are described in Stansberry et al. (2008) and Brucker et al. (2009). An alternate, simpler method, to obtain final maps is to co-add directly the original single maps (i.e. not background-corrected) in the target frame. This method is obviously less optimal than the previous one in terms of SNR, but provides a useful test to demonstrate that the background subtraction is not introducing any spurious effects. A more complete and detailed description about the data reduction process can be found in Appendix A, and will be published in Kiss et al. (\textit{in prep.}-a). 
   
   
\subsection{Photometry}
\label{photom}

We use IRAF/Daophot\footnote{Image Reduction and Analysis Facility (IRAF), is a collection 
of software written at the National Optical Astronomy Observatory (NOAO) geared towards 
the reduction of astronomical images in pixel array form.} to perform standard synthetic-aperture photometry on
the final sky-subtracted, co-added maps for our targets. The target is
typically obvious and located very near the center of the maps. For very
faint targets and/or ones with some residual structure from the background
subtraction (as can happen in the green and red channels), the ephemeris
position of the target is used to position the photometric aperture. 

Once the target is identified, we measure the flux at the photocenter position for aperture radii ranging from 1 to 15 pixels. We perform an aperture correction technique (\cite{1989PASP..101..616H}) for each aperture radius using the encircled energy fraction for a point--source for the PACS instrument\footnote{M\"uller et al. 2011: PACS Photometer -Point-Source Flux Calibration, PICC-ME-TN-037, Version 1.0; Retrieved November 23, 2011; \url{http://herschel.esac.esa.int/twiki/pub/Public/PacsCalibrationWeb/pacs_bolo_fluxcal_report_v1.pdf}}. We construct an aperture-corrected curve-of-growth using these fluxes (red line of Figure \ref{Typhon}). Given the optimized reduction technique above, the fluxes used to construct these aperture-corrected curves-of-growth need not be corrected for background contribution.

After inspection of the curves-of-growth, the optimum synthetic aperture to measure the target flux is selected, and is typically $\sim$ 1.0-1.25 times the point spread function (PSF) FWHM in radius (PSF FWHM in radius is $5.2\arcsec/7.7\arcsec/12\arcsec$ in blue/green/red bands respectively). Usually,  the optimum aperture radius lies in the ``plateau" zone of the curves-of-growth (see Figure \ref{Typhon}). In cases in which there are other sources or artifacts very close to the target, we use a smaller aperture radius to avoid contamination of the target flux.

Uncertainties on the flux measurements are estimated by means of a Monte-Carlo technique, in which 200 artificial sources are implanted on the actual
background-subtracted final maps. These artificial sources have the structure of the PSF for the specific (blue, green, or red) channel. Sources are randomly implanted on a square region of $50\arcsec \times 50\arcsec$ around the target photocenter, with an exclusion zone corresponding  
to a circular region around the target photocenter with a radius $= 2\times$PSF FWHM. We measure and aperture-correct the fluxes of these artificial sources using an aperture radius of 5 re-sampled pixels, which is the median optimum aperture radius. Uncertainties are computed as the standard deviation of these 200 fluxes, and finally
multiplied by a factor $\sqrt 2$. The latter factor is due to the fact that given the way the final maps are generated, the remaining background is measured only once 
in the immediate vicinity of the real target but two times in the rest of the image (i.e. for the artificial sources). Note that except for the background subtraction, all the above reduction and photometric steps were validated on standard stars with known flux.

The fluxes and associated uncertainties are finally color-corrected for each band. Flux densities within the PACS photometric system are defined as those that a source with a flat spectrum would have at the PACS reference wavelengths of 70, 100 and 160 $\mu m$ (Poglitsch et al. 2010). Converting the in-band flux to monochromatic flux densities therefore depends on the spectral energy distribution of the source and the transmission profile for each PACS band. 
Color-corrections for each filter are tabulated as a function of color temperature\footnote{M\"uller et al. 2011, PACS Photometer Passbands and Colour Correction Factors for Various Source SEDs, PICC-ME-TN-038, Version 1.0; Retrieved November 23, 2011; \url{http://herschel.esac.esa.int/twiki/pub/Public/PacsCalibrationWeb/cc_report_v1.pdf}}  but are usually within a few percent except at the very low end ($T$ $<$ 30 K) of the temperature range. For the color temperature, we used an approximation of the disk-averaged temperature, i.e. T$_{color}$ = T$_{ss} / 2^{1/4}$, where T$_{ss}$, the subsolar temperature (see equation (\ref{eq_tss}) ), was obtained from a first run of the thermal model described below, using at this step the in-band (i.e. color-uncorrected) fluxes. T$_{color}$ values for our sample range from 25 K for Eris to 77 K for 2002 PN$_{34}$. The mean color correction factors for all our sample are: 0.991$\pm$0.028 (blue band), 0.989$\pm$0.006 (green band), and 1.011$\pm$0.017 (red band). The impact of these factors is smaller than the uncertainties of our measurements.

The absolute flux calibration of PACS is based on standard stars and large main belt asteroids and has uncertainties of 3\% for the blue/green channel, and 5\% for the red channel\footnote{M\"uller et al. 2011: PACS Photometer -Point-Source Flux Calibration, PICC-ME-TN-037, Version 1.0; Retrieved November 23, 2011; \url{http://herschel.esac.esa.int/twiki/pub/Public/PacsCalibrationWeb/pacs_bolo_fluxcal_report_v1.pdf}}. The final flux uncertainties we use in our modeling are the root-sum-square
combination of these absolute calibration uncertainties and the uncertainties
we obtain from the photometric measurements for individual targets. A summary of the final, color-corrected and absolute-calibrated, fluxes of our SDOs/Detached sample is given in Table~\ref{Results}.

\begin{table}

\caption{Color corrected fluxes at 70, 100 and 160 $\mu m$ for the SDOs/Detached sample} 

\label{Results} 

\centering 

\scalebox{0.85}{
\begin{tabular}{r c c c c} 

\hline\hline 

Object	&	Flux$_{70}$[mJy]	&	Flux$_{100}$[mJy]	&	Flux$_{160}$[mJy]	&	Class.	\\


\hline 

(15874) 1996 TL$_{66}$	&	9.6$\pm$1.2	&	9.7$\pm$1.1	&	9.2$\pm$2.2	&	SDO	\\%
(82158) 2001 FP$_{185}$	&	9.2$\pm$0.8	&	11.1$\pm$1.1	&	7.0$\pm$1.9	&	SDO	\\%
(73480) 2002 PN$_{34}$	&	11.4$\pm$1.2	&	11.9$\pm$1.4	&	7.6$\pm$6.1	&	SDO	\\%
(127546) 2002 XU$_{93}$	&	16.5$\pm$0.9	&	15.0$\pm$1.6	&	6.2$\pm$2.1	&	SDO	\\%
2007 OR$_{10}$	&	3.7$\pm$1.1	&	4.8$\pm$1.4	&	6.7$\pm$2.3	&	SDO	\\%
2007 RW$_{10}$	&	13.2$\pm$0.9	&	11.3$\pm$1.3	&	10.3$\pm$2.1	&	SDO	\\%
(65489) Ceto	&	12.3$\pm$0.7	&	10.5$\pm$1.0	&	6.2$\pm$4.8	&	SDO	\\%
(42355) Typhon	&	28.5$\pm$1.2	&	21.4$\pm$1.2	&	9.1$\pm$2.4	&	SDO	\\%
(40314) 1999 KR$_{16}$	&	5.7$\pm$0.7	&	3.5$\pm$1.0	&	4.6$\pm$2.2	&	DO	\\%
(120132) 2003 FY$_{128}$	&	14.4$\pm$0.9	&	14.0$\pm$1.2	&	13.0$\pm$2.2	&	DO	\\%
2005 QU$_{182}$	&	4.5$\pm$0.9	&	2.5$\pm$1.1	&	8.4$\pm$3.0	&	DO	\\%
(145480) 2005 TB$_{190}$	&	5.8$\pm$0.9	&	8.6$\pm$1.4	&	5.4$\pm$1.8	&	DO	\\%
2007 OC$_{10}$	&	8.3$\pm$0.9	&	7.8$\pm$1.2	&	7.4$\pm$1.8	&	DO	\\%
(229762) 2007 UK$_{126}$	&	13.7$\pm$1.1	&	12.3$\pm$1.3	&	7.2$\pm$2.2	&	DO	\\%
(136199) Eris	&	2.0$\pm$0.6	&	3.7$\pm$1.0	&	5.6$\pm$1.6	&	DO	\\%

\hline 

\end{tabular}}

\footnotesize{
 \textbf{Flux$_{70}$, Flux$_{100}$, Flux$_{160}$:} Color-corrected fluxes (mJy) at 70, 100 and 160 $\mu m$. 
\textbf{Class.}, dynamical classification following Gladman et al. 2008. SDO = scattered disc object, DO = detached object.}

\end{table}

\subsection{\textit{Spitzer}-MIPS observations}
\label{spitzer-mips}

About 102 TNOs/Centaurs have been observed with the 0.85 m \textit{Spitzer} Space Telescope (\cite{2004ApJS..154....1W}) using the Multiband Imaging Photometer (MIPS) instrument (\cite{2004ApJS..154...25R}). The MIPS 24 $\mu m$ channel consist of a $128\times128$ pixel detector and the 70 $\mu m$ channel of a $32\times32$ pixel detector. For these 2 channels the telescope-limited resolution is 6$\arcsec$ and 18$\arcsec$ respectively. Fluxes for 60 objects have been published in Stansberry et al. (2008) and Brucker et al. (2009) but others are unpublished. Eight targets of our sample have available (published or unpublished) \textit{Spitzer}-MIPS flux density measurements at 24 and/or at 70 $\mu m$ (see Table \ref{MIPS_tabla}). Unpublished data have been processed in a similar way as in Stansberry et al. (2008) and Brucker et al. (2009), using the calibration (\cite{Gordon2007}, \cite{Engelbracht2007}) and color-corrections (\cite{stansberry07}) described in the references. As in Stansberry et al. (2008), we adopt absolute calibration uncertainties of 3\% and 6\% for the 24 and 70 $\mu m$ observations (50\% larger than the uncertainties derived for observations of stellar calibrators). The larger uncertainties allow for effects from the background-subtraction technique, the faintness of the targets, and uncertainties in the color corrections. The background subtraction techniques used for our \emph{Herschel} data (see Sect. \ref{DataRed}) are in fact derived from techniques originally developed for the MIPS data reductions. The previously unpublished fluxes presented here are based on new reductions of the MIPS data, utilizing updated ephemeris positions for the targets. This allows for more precise masking of the target when generating the image of the sky background, and for more precise placement of the photometric aperture. In most cases the new fluxes are very similar to the previously published values for any given target, but in a few cases significant improvements in the measured flux and SNR were achieved.

As discussed below, the combination of \textit{Spitzer}-MIPS and \textit{Herschel}-PACS fluxes significantly improves results from the thermal modeling. The \emph{Spitzer} 70 $\mu m$ band is similar to the \emph{Herschel} 70 $\mu m$ channel, providing a consistency check on data from the two observatories. The inclusion of the \textit{Spitzer} 24 $\mu m$ flux, when available, strongly improves the ability to constrain the surface temperature distribution, i.e. in general, allows us to determine the beaming factor (see Sect. \ref{thermalmodel}). Published, re-analyzed, and unpublished \textit{Spitzer}-MIPS flux density measurements available for our sample are given in Table \ref{MIPS_tabla}. Note that the effective wavelengths for these fluxes are 23.68 and 71.42 $\mu m$.

\begin{table*}

\caption{\textit{Spitzer}-MIPS color corrected fluxes at 23.68 and 71.42 $\mu m$ available for our SDOs/Detached sample.} 

\label{MIPS_tabla} 

\centering 

\begin{tabular}{r c c c c c c} 

\hline\hline 

Object	&	Mid-time &$r_h$[AU] & $\Delta$ [AU] & $\alpha$[deg] & Flux$_{24}$[mJy]	&	Flux$_{70}$[mJy] \\


\hline 

(15874) 1996 TL$_{66}$	& 2008-Feb-17 04:27:11 & 35.4370 & 34.9378 & 1.4 & 0.324$\pm$0.020$^{2}$ &	10.12$\pm$1.40$^{2}$	\\
(82158) 2001 FP$_{185}$	& 2004-Jun-20 22:51:52 & 34.2594 & 34.0927 & 1.7 & 0.395$\pm$0.096$^{3}$	&	Contaminated flux	\\
(73480) 2002 PN$_{34}$	& 2006-Jul-16 11:00:56 & 14.6074 & 14.1524 & 3.6 & 10.144$\pm$0.117$^{2}$	&	30.34$\pm$1.59$^{2}$	\\
(127546) 2002 XU$_{93}$	& 2006-Nov-04 20:09:23 & 21.0896 & 20.8251 & 2.7 & 2.431$\pm$0.052$^{3}$	&	14.12$\pm$1.19$^{3}$	\\
(65489) Ceto	& 2006-Jul-19 03:19:41 & 27.9924 & 27.6753 & 2.0 & 1.419$\pm$0.047$^{2}$	&	14.6$\pm$2.6$^{1}$	\\
(42355) Typhon	& 2008-Jun-26 22:18:45 & 17.6784 & 17.5110 & 3.3 & 4.974$\pm$0.125$^{3}$	&	27.46$\pm$3.43$^{2}$	\\
(120132) 2003 FY$_{128}$& 2007-Jul-14 07:30:00 & 38.3556 & 38.1281 & 1.5 & 0.427$\pm$0.051$^{3}$ &	19.38$\pm$3.16$^{3}$ \\
(136199) Eris	& 2005-Aug-24 22:13:07 & 96.9049 & 96.4083 & 0.5 & $<0.005^{1}$	&	2.7$\pm$0.7$^{1}$	\\

\hline 

\end{tabular}

\begin{flushleft}
\footnotesize{
\textbf{Mid-time:} Mean date and UT time of the observation. 
\textbf{r [AU]:} heliocentric distance (AU) for the observation date. 
\textbf{$\Delta$ [AU]:} distance object-\textit{Spitzer} (AU) for this date. 
\textbf{$\alpha$[deg]:} phase angle -degrees.
\textbf{Flux$_{24}$, Flux$_{70}$:} \textit{Spitzer}-MIPS color-corrected flux densities (mJy) at 23.68 $\mu m$ and 71.42 $\mu m$. Upper limits are 1$\sigma$. 
\textbf{References:} $^{1)}$ \cite{stansberry08}; $^{2)}$ Revised data from \cite{stansberry08}; $^{3)}$ Previously unpublished data.}
\end{flushleft}

\end{table*}


\section{Thermal model}
\label{thermalmodel}

\subsection{General formulation}
The combination of thermal-infrared and optical data makes it possible to infer an objects' diameter ($D$), 
geometric albedo ($p_V$) and constrain its thermal properties. This requires the use of a thermal model 
to calculate the disk-integrated thermal emission. Here we use the so-called hybrid standard 
thermal model (hybrid STM, M\"uller et al. 2010, Stansberry et al. 2008). Except for the fact that the hybrid STM assumes 
zero phase angle (which is an excellent approximation for TNOs), it is identical to the Near-Earth Asteroid 
Thermal Model (NEATM, \cite{Harris98}). NEATM was originally developed for Near Earth Objects (NEOs) but is applicable to all atmosphereless 
bodies and is being used in companion papers by Mommert et al. (2012) and Vilenius et al. (2012). In this model, the temperature distribution follows instantaneous equilibrium of a smooth
spherical surface with solar input, but is modified
by a factor $\eta$ (beaming factor), which represents empirically the combined effects of
thermal inertia and surface roughness. As shown by Harris (1998) and Stansberry et al. (2008), multi-wavelength
measurements can be used to constrain $\eta$ if the data are of sufficient
quality and encompass both the short-wavelength and the peak and/or long-wavelength 
portions of the spectral energy distribution (SED) of the target.
In this case the data give a direct measurement of the color-temperature of
the emission, which is sensitive to the value of $\eta$. 
Our preferred approach was to simultaneously model \textit{Herschel} and \textit{Spitzer} measurements (of course accounting
for their different observer-centric, $\Delta$, and heliocentric, $r_h$, distances) and fit an $\eta$ factor to
the combined dataset. We call this the ``floating--$\eta$" approach. When \textit{Spitzer} data were unavailable (or when this method led to implausible $\eta$ values; see \cite{Mommert2011}) we instead
used a ``fixed--$\eta$" approach, with $\eta = 1.20 \pm 0.35$, the mean value derived by \cite{stansberry08} from \textit{Spitzer} observations of TNOs.

The model therefore fits the \textit{Herschel} (3 data points) or \textit{Herschel} + \textit{Spitzer} (5 data points) fluxes 
in terms of two ($D$ and $p_V$) or three ($D$, $p_V$, and $\eta$) parameters. The model also requires the choice
on a phase integral $q$, that we assumed to be related to $p_V$ through
$q = 0.336 \cdot p_V +0.479$ (\cite{Brucker2009}). Diameter and geometric albedo are further related 
through: 

\begin{equation}
\label{eq_diam}
	   	D = 2 \cdot 10^{V_{\sun}/5} \cdot 10^{-H/5}/\sqrt{p_V} \cdot 1 \mbox{AU/km}  
\end{equation} where $H$ is the absolute magnitude (nominally in V band), and $V_{\sun} = -26.76\pm0.02$ (\cite{Bessell1998}) the V magnitude of the Sun. A grey emissivity ($\epsilon$= 0.9) was assumed when calculating the
local temperatures and monochromatic fluxes. With these definitions the subsolar temperature is given by: 
\begin{equation}
\label{eq_tss}
	T_{ss}=  \left( \frac{F_{\sun} \cdot (1-p_V \cdot q)} {r_h^2 \cdot \eta \cdot \epsilon} \right) ^{1/4} 
\end{equation}where F$_{\sun}$ is the solar flux at 1 AU. The solution for the $D$, $p_{V}$ (and $\eta$) parameters
are determined from a $\chi^2$ minimization of the deviation of the model from the observed fluxes, accounting for their individual
1-$\sigma$ error bars. Note that when only upper limits were available, 
they were treated as non-detections (i.e. zero flux) with a 1-$\sigma$ uncertainty as the upper limit. In the case of multiple systems $D$ is the area-equivalent diameter, i.e., that of the sphere with the same total surface area, because none of the binaries are resolved in our data.

\subsection{H magnitudes}
\label{Hmag_issue}

Equation (\ref{eq_diam}) above outlines the role of absolute magnitude ($H$) in the thermal modeling. Determining accurate values of $H$, preferably in $V$ band, is therefore an essential task for this project. We describe hereafter our adopted approaches.

In the best situation, photometrically calibrated observations are available at a variety of phase angles. In this case, correcting the
observed $V$ magnitudes for the heliocentric ($r_h$) and geocentric ($\Delta$) distances provides $m_{V}(1,1)= V - 5\cdot log(r\cdot \Delta)$. A linear fit of the $m_{V}(1,1)$ magnitudes vs phase angle $\alpha$  ($m(1,1)= H_V + \beta\cdot \alpha$), taking into consideration measurement
errors, then provides the phase coefficient $\beta$ (mag/deg) and the absolute magnitude $H_V$, and their associated error bars. Whenever data are unsufficient in $V$ band but
adequate in R band, we similarly and alternatively infer $H_R$, which by means of the thermal model will yield the red geometric albedo ($p_{R}$).

In the case when not enough data in either V or R  are available, 
or if the above approach fails due to poor quality data, we adopt a mean $\beta= 0.10\pm0.04$ mag/deg which is the average value (excluding Pluto) from \cite{Belskaya2008}.
We then derive H magnitudes for each individual photometric measurement and use the weighted average result.

When photometrically-calibrated magnitudes are not available in the literature, we use astrometric observations compiled by MPC, as those include rough estimates of the optical magnitudes (most often in $R$ band). The above approaches are then applied to this data, bearing in mind that individual V or R magnitudes reported in MPC data may have large (and systematically biased) uncertainties up to 0.5 magnitudes (\cite{2005Icar..179..523R}). 

When full lightcurve information is available (i.e. the amplitude is known and the lightcurve can be temporally phased to the
\textit{Herschel} measurements), the $H$ magnitude is corrected accordingly.  
When the lightcurve amplitude (or an upper limit) is known, but the period is unknown or phasing is impossible, 88\% of half peak-to-peak amplitude (i.e. 1$\sigma=$  68\% of the values of a sinusoid are within this range) is quadratically added to the uncertainty on $H$. Finally, if there is no lightcurve information, we assume peak-to-peak amplitude less than 0.2 mag, as based on Duffard et al. (2009), $\sim$70\% of the objects have a peak-to-peak amplitude less than this. In this case we quadratically add 0.09 mag (i.e. 88\% of 0.2/2) to the $H$ uncertainty. Absolute magnitudes quoted in Table \ref{tableOBS} have been obtained as just described.

Note finally that in case we have to use R magnitudes (i.e. for 2007 OR$_{10}$, 2007 RW$_{10}$ and 1999 KR$_{16}$) modeling provides the R geometric albedo ($p_R$) instead of V albedo ($p_V$). $p_{R}$ can be converted to $p_{V}$ through:

\begin{equation}
\label{albedoconversion}
  p_{V}= p_{R}\cdot 10^{[((V-R)_{\odot}-(V-R)_{obj}))/2.5]}
\end{equation} 
where $(V-R)_{\odot}$= 0.36 mag is the Sun's V-R, and $(V-R)_{obj}$ that of the object. 
We assume $(V-R)_{obj}$= 0.55$\pm$0.11, the mean value from Hainaut \& Delsanti (2002) for 66 SDOs/Detached objects, except for 2007 OR$_{10}$ (see discussion for this object in Sect. \ref{large_objects}).

\begin{figure*}[!hpbt]
   \centering

   \includegraphics[width=4.4cm,angle=90]{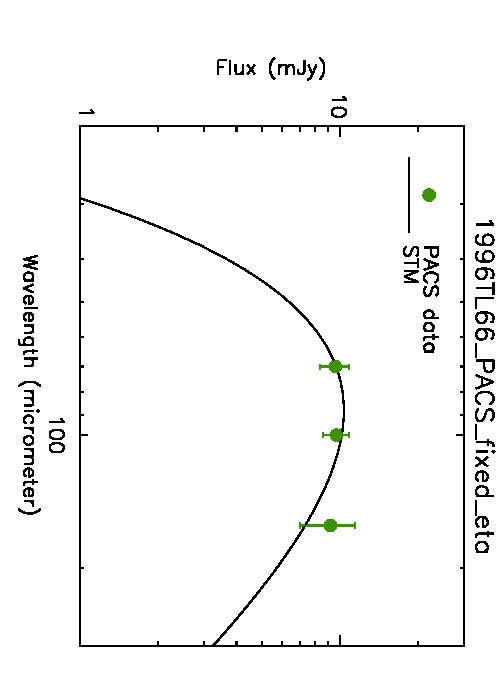}
   \includegraphics[width=4.4cm,angle=90]{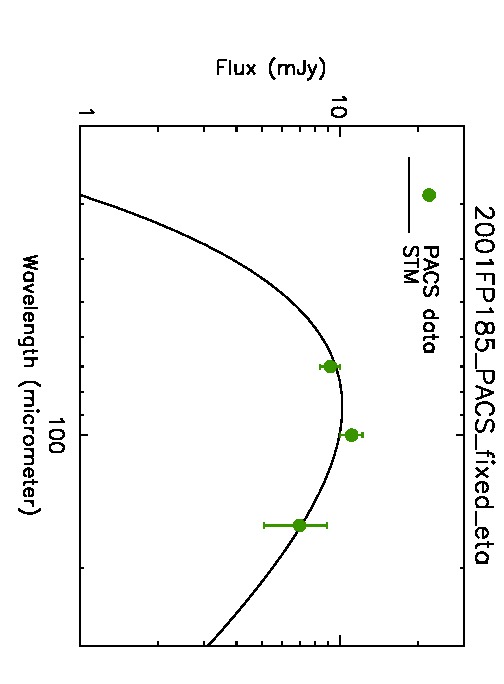}
   \includegraphics[width=4.4cm,angle=90]{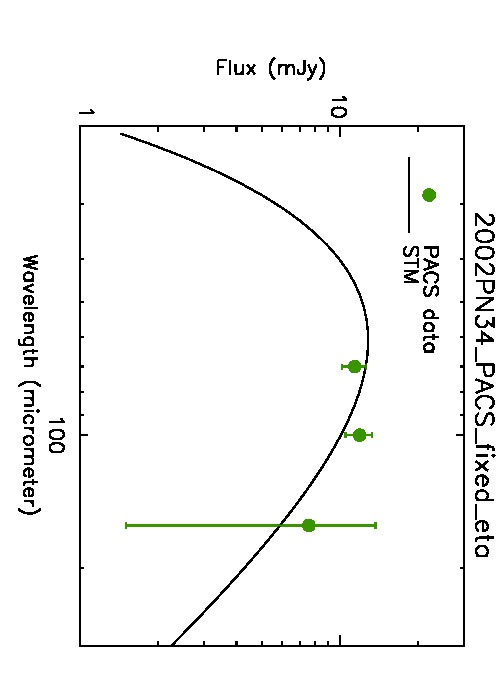}
   \includegraphics[width=4.4cm,angle=90]{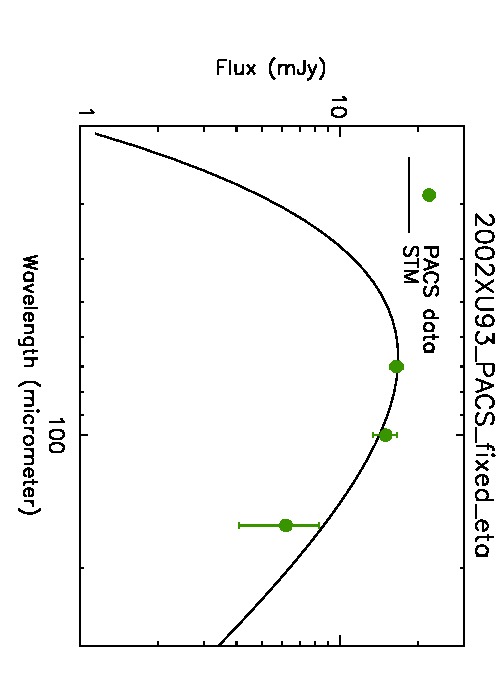}
   \includegraphics[width=4.4cm,angle=90]{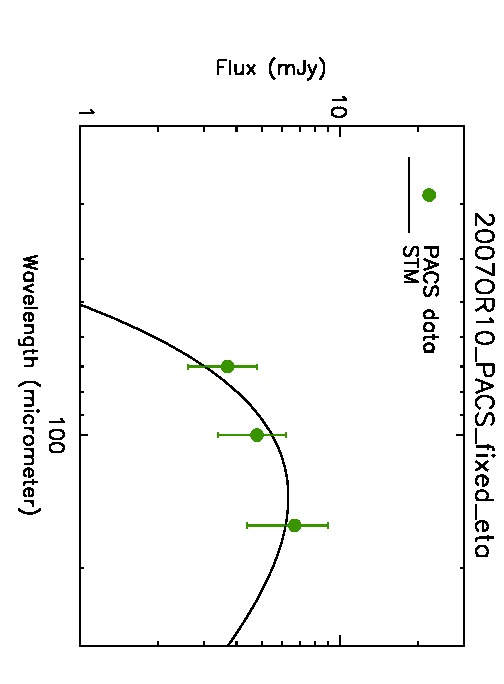}
   \includegraphics[width=4.4cm,angle=90]{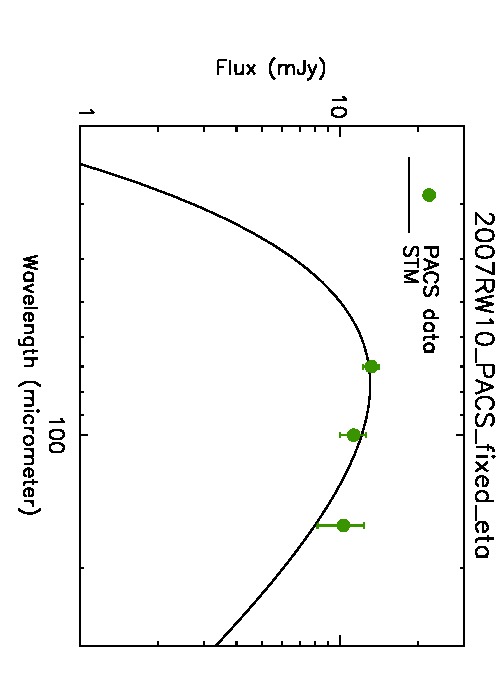}
   \includegraphics[width=4.4cm,angle=90]{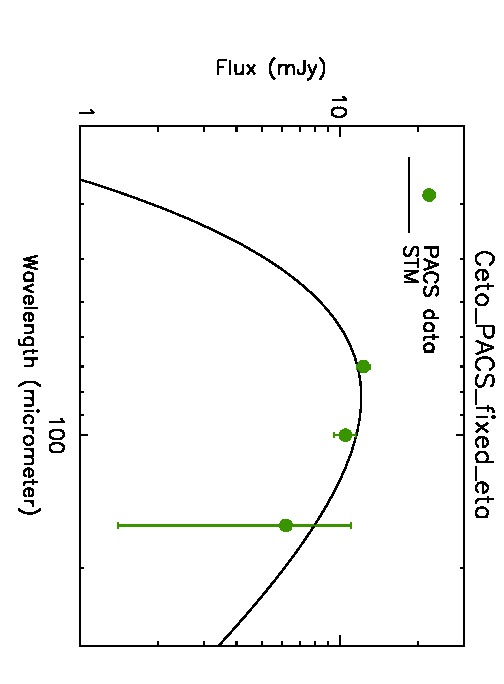}
   \includegraphics[width=4.4cm,angle=90]{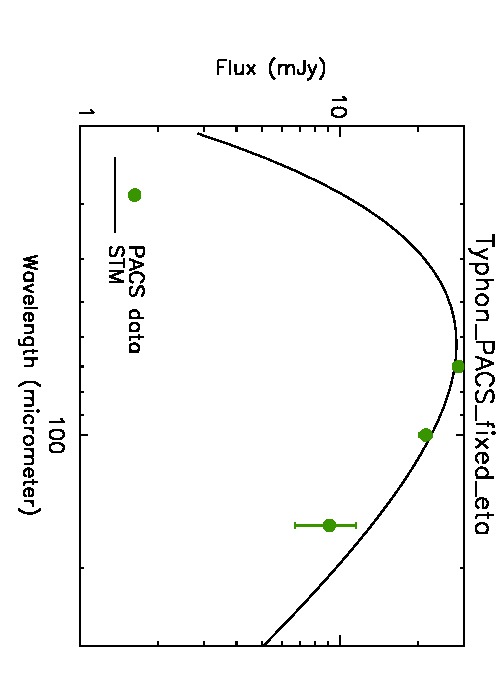}
   \includegraphics[width=4.4cm,angle=90]{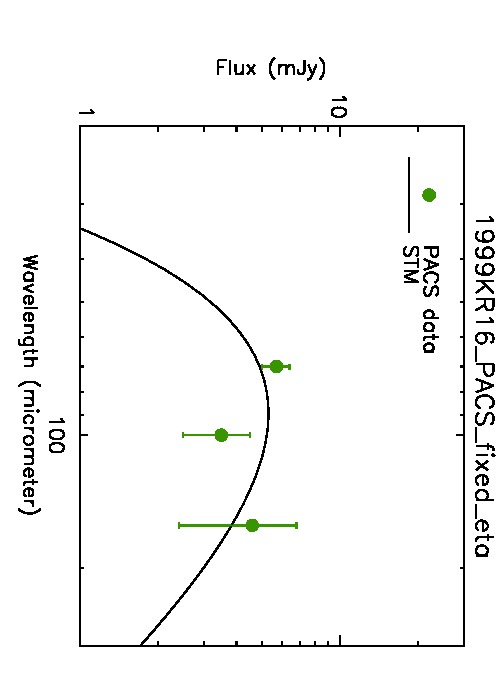}
   \includegraphics[width=4.4cm,angle=90]{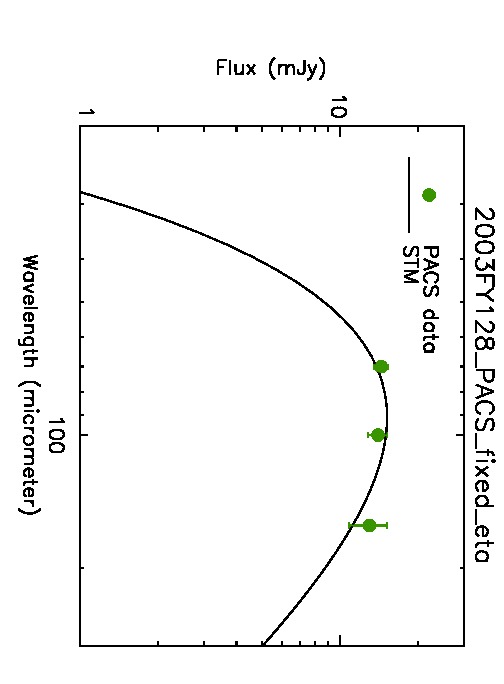}
   \includegraphics[width=4.4cm,angle=90]{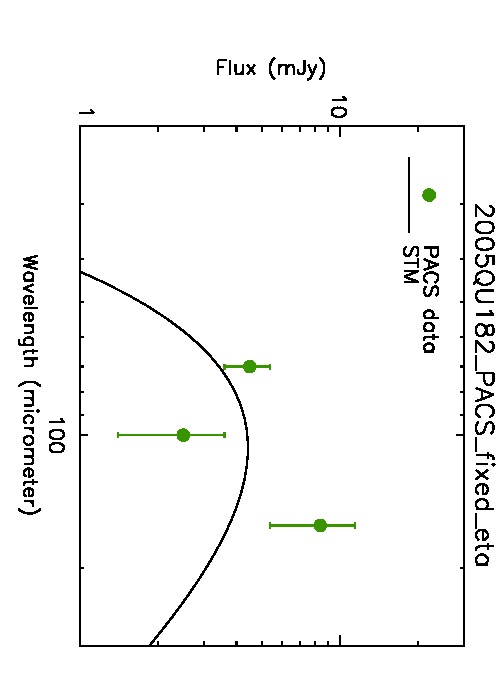}
   \includegraphics[width=4.4cm,angle=90]{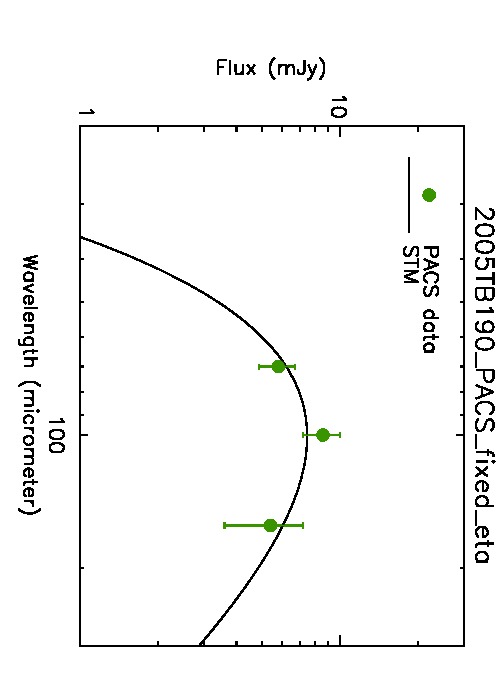}
   \includegraphics[width=4.4cm,angle=90]{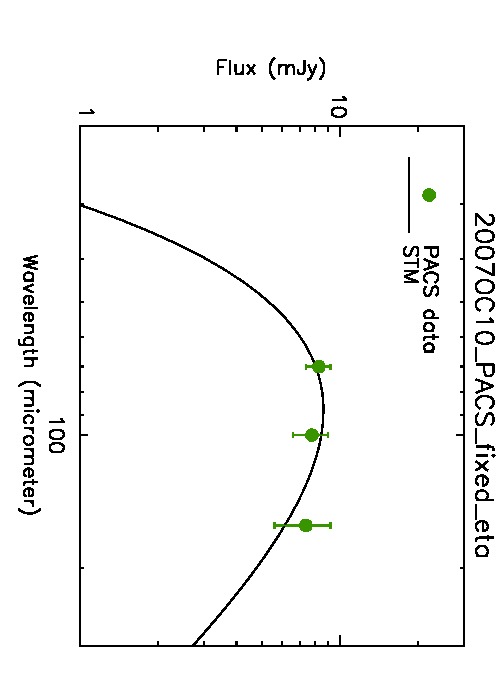}
   \includegraphics[width=4.4cm,angle=90]{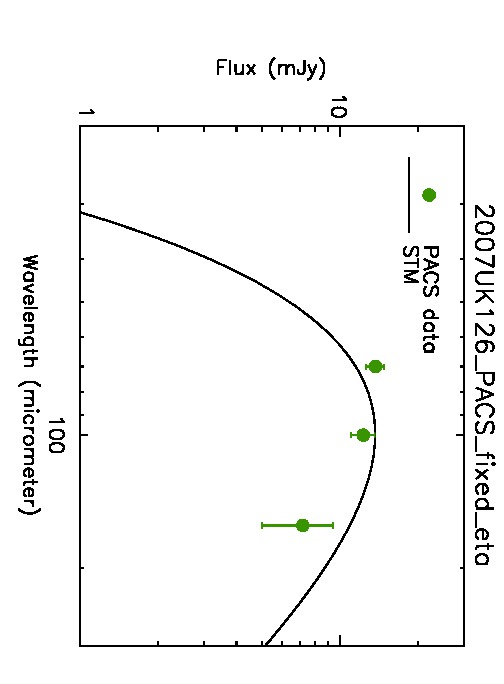}
   \includegraphics[width=4.4cm,angle=90]{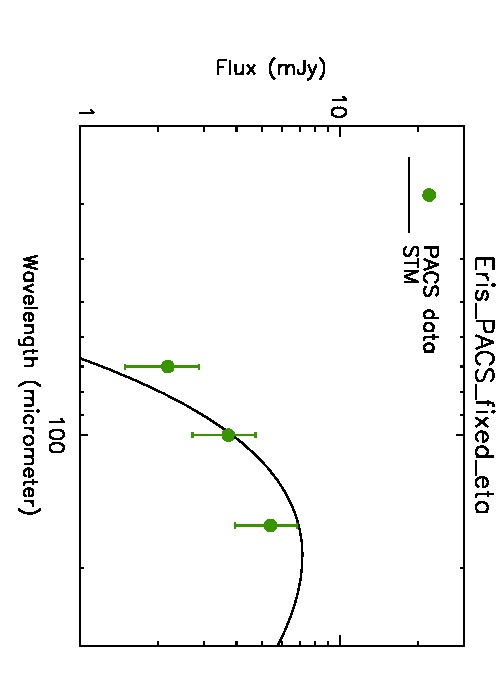}
   
      \caption{Radiometric fixed-$\eta$ fits for the objects observed only with \textit{Herschel}-PACS.}
              
         \label{PACS_fits}
   \end{figure*}

\begin{figure*}[!hpbt]
   \centering
   
   \includegraphics[width=4.4cm,angle=90]{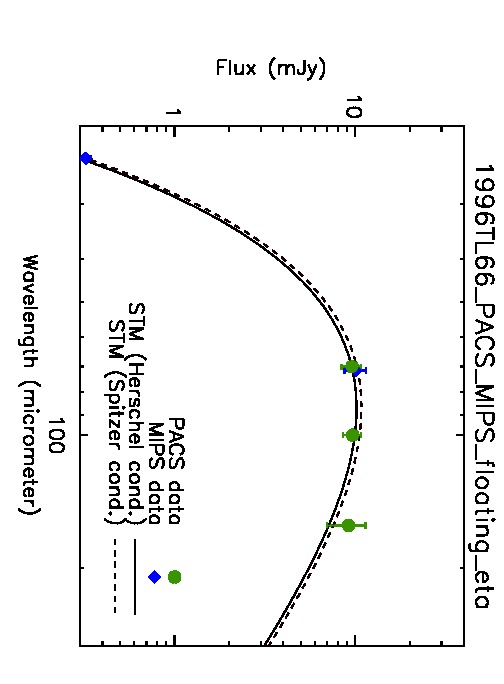}
   \includegraphics[width=4.4cm,angle=90]{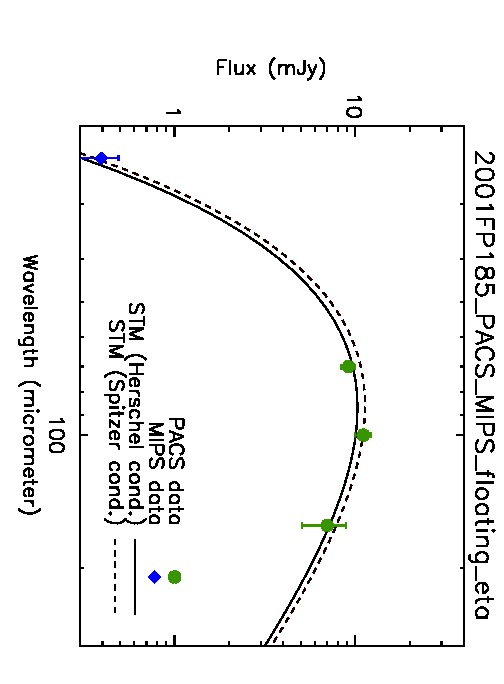}
   \includegraphics[width=4.4cm,angle=90]{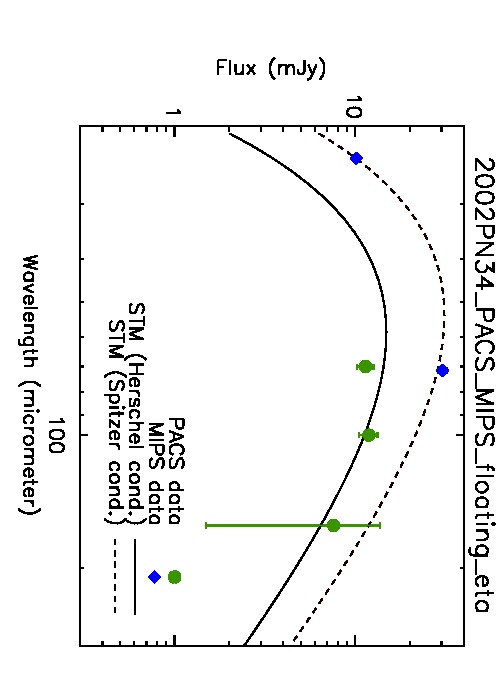}
   \includegraphics[width=4.4cm,angle=90]{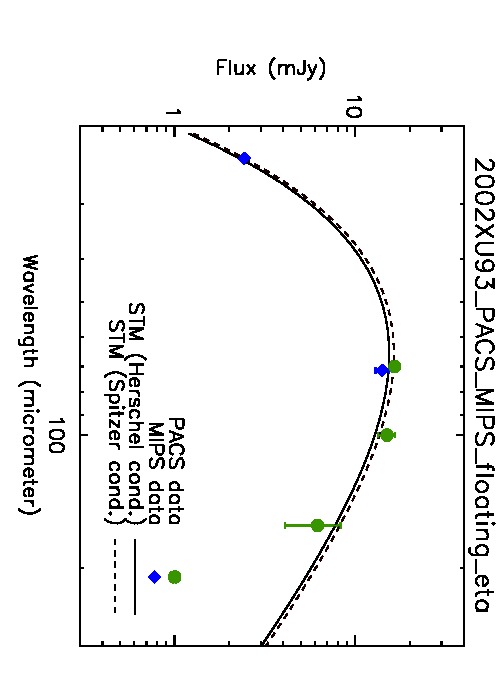}
   \includegraphics[width=4.4cm,angle=90]{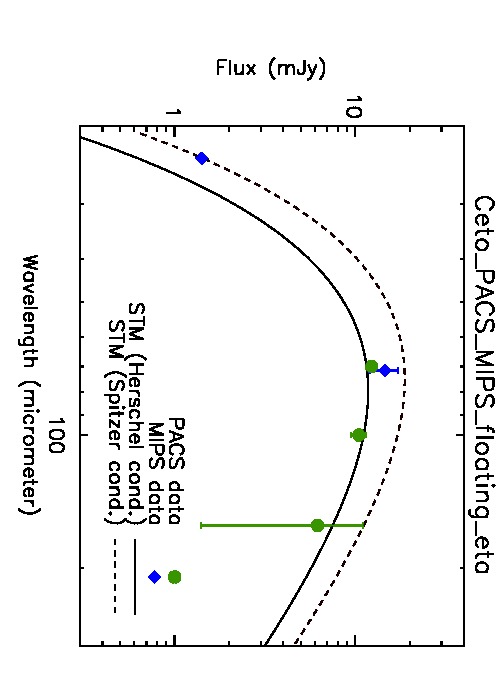}
   \includegraphics[width=4.4cm,angle=90]{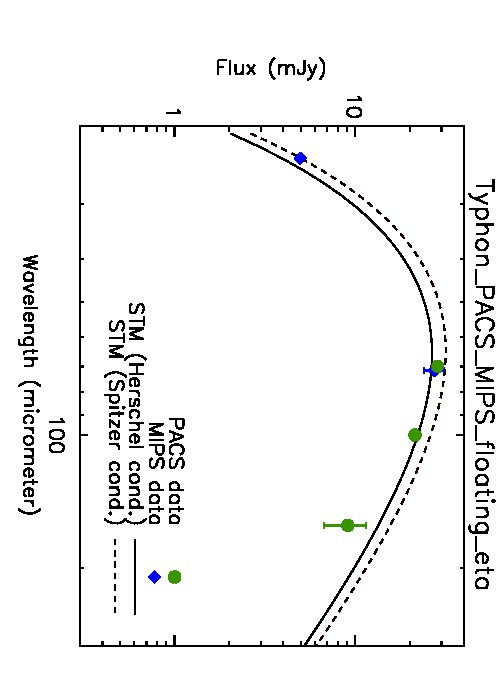}
   \includegraphics[width=4.4cm,angle=90]{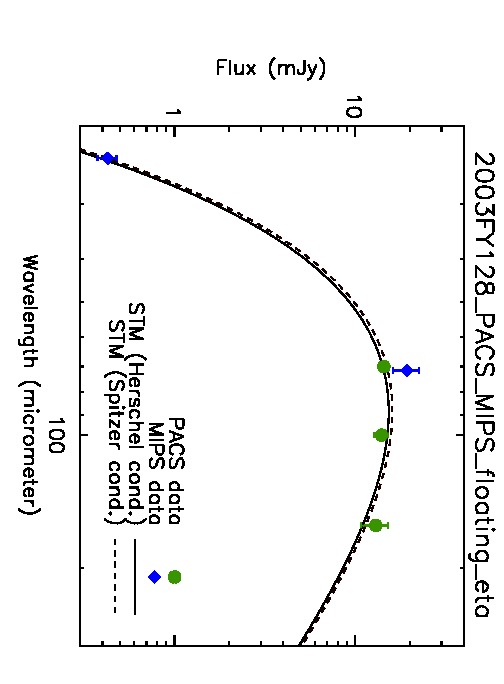}
   \includegraphics[width=4.4cm,angle=90]{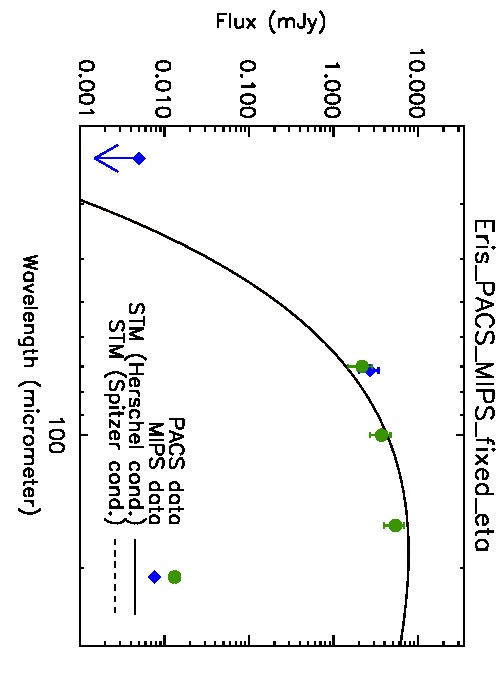}
         
      \caption{Radiometric hybrid-STM fits for the objects observed with \textit{Spitzer}-MIPS and \textit{Herschel}-PACS. The best fit is plotted for the \textit{Herschel} ($r_h$, $\Delta$) conditions (solid line) and for the \textit{Spitzer} conditions (dashed line).
              }
         \label{MIPS_PACS_fits}
   \end{figure*}

\subsection{Uncertainties}
\label{MCapproach}

Within the framework of the thermal model, uncertainties on the fitted ($D$, $p_V$, $\eta$) parameters were obtained using a Monte Carlo approach (see Mueller
et al. 2011) in which 1000 synthetic datasets were randomly generated using the 
uncertainties in the measured fluxes. Note that the uncertainty in the $H$ magnitude (and the $\pm$0.35
uncertainty in $\eta$, in the case of the ``fixed-$\eta$" approach) was also accounted for in 
the Monte-Carlo approach. Naturally, we eliminated those Monte-Carlo simulations which generated
negative fluxes (or negative $\eta$). The diameter,
albedo (and $\eta$) of each of the 1000 synthetic objects were determined, and their distributions were used to
determine the error bars on these parameters. As outlined in Mueller
et al. (2011), these distributions are generally not Gaussian (especially the albedo distribution).
Therefore, we adopted the median of the Monte-Carlo results as the nominal value, and
asymmetric error bars to include the 68.2 \% of the results. 

Figure \ref{PACS_fits} shows best fits for all objects in our sample using only the \textit{Herschel}-PACS measurements. All these fits
are ``fixed-$\eta$". For the 8 objects that also have \textit{Spitzer} fluxes, Figure \ref{MIPS_PACS_fits} shows best fits (``floating-$\eta$",
except for Eris, which is not detected at 24 $\mu$m) for the combined
\textit{Herschel} and \textit{Spitzer} measurements. In that Figure, the same best fit model is plotted twice, once (solid line) for
the \textit{Herschel} ($r_h$, $\Delta$) conditions and once for the \textit{Spitzer} (dashed line) conditions.
In general, this shows that the model appropriately accounts for rather different \textit{Herschel} vs. \textit{Spitzer} 
fluxes associated to different $r_h$ and $\Delta$ (e.g. Ceto, 2001 FP$_{185}$). Many of the fits shown
in Figure \ref{PACS_fits} and \ref{MIPS_PACS_fits} are satisfactory, in that the model fits all data points within or close to 
the 1-$\sigma$ error bars. However, in a number of cases, the fits are obviously bad (e.g. 1999 KR$_{16}$, 2005 QU$_{182}$, Typhon); in those cases, the data points are mutually inconsistent and the model
is no more than a mere compromise between them. Given that the PACS green vs blue observations (and the
\textit{Spitzer} vs \textit{Herschel} data) are not taken simultaneously, such behaviour could in principle result from 
rotational variability of the objects. An alternate physical explanation would be that the fluxes
are affected by spectral features, e.g. depressing the 100-$\mu$m flux with respect to the 70-$\mu$m
and 160-$\mu$m values, or more generally that our simple STM-like approach
does not capture the complexity of the thermal emission. However, the discrepancies are often large and these explanations
may not be quantitatively plausible. In those cases, we conclude that the error bars we have determined
on the fluxes are probably underestimates of the true error bars, though we are unable to track down the
reasons for that. 

In the situation where the formal measurement errors are too small (or similarly that the model
is not completely adequate), the above described Monte-Carlo method will
underestimate the uncertainties on the solution parameters. To handle this problem, 
we adopted a ``rescaled error bar" approach described in detail in Appendix B (Sect. \ref{infladas}). Essentially,
it consists of ``inflating"  the measurement error bars by a suitable factor before running the Monte-Carlo
simulation. This approach was applied to
the six objects for which the radiometric fit is poor (see Figs. \ref{PACS_fits} and \ref{MIPS_PACS_fits}): 
1999 KR$_{16}$, 2002 PN$_{34}$, 2002 XU$_{93}$, 2005 QU$_{182}$, 2007 UK$_{126}$, and Typhon, 
with rescaling factors varying from 1.45 to 1.99. Fluxes for the remaining nine objects are satisfactorily fit 
within the 1-$\sigma$ flux uncertainties and did not require any adjustment of the measurement errors. We finally stress that the
diameter, albedo, and beaming factor values and their uncertainties
reported in the next section are obtained in the framework of the hybrid
STM, which may have its own limitations (see Harris 2006 for a discussion
in the case of NEOs). More physical models (e.g. TPM) may lead to somewhat
different values and error bars, as illustrated by M\"uller et al. (2010) in
the case of a few objects observed in the Science Demonstration Phase (SDP) of the ``TNOs are
cool" programme.


\section{Results and discussion}
\label{results_and_discussion}


\subsection{General results}
\label{generalresults}

Results of the model fits in terms of the objects diameter, albedo and beaming factor are gathered in Table 5. When \textit{Spitzer}-MIPS measurements
are available (8 objects), the model was run twice, once considering the PACS fluxes only, and once combining PACS and MIPS.
As illustrated in Table 5, these two fitting options give consistent solutions within error bars. The PACS+MIPS solution
is always adopted as the preferred one (outlined in bold face in Table \ref{RadiometricResults}), as it (i) provides reduced error bars (ii) permits the determination of the beaming factor.

Table  \ref{RadiometricResults} includes first size/albedo determination for 9 objects (4 SDO and 5 detached objects). Six of them
are based on PACS-only measurements (2007 OR$_{10}$, 2007 RW$_{10}$, 1999 KR$_{16}$, 2005 QU$_{182}$, 2007 OC$_{10}$, and 2007 UK$_{126}$), and the
remaining three make use of PACS plus unpublished-MIPS data (2001 FP$_{185}$, 2002 XU$_{93}$, and 2003 FY$_{128}$). 
Published sizes and albedos exist for the other 6 objects of our sample (1996 TL$_{66}$, 2002 PN$_{34}$, Ceto, Typhon, 2005 TB$_{190}$ and Eris) but we here present
fluxes for these objects at wavelengths not observed before (i.e. 100 $\mu m$ and 160 $\mu m$), and the combination with earlier measurements leads to improved
estimations of sizes, albedos, and beaming factors (see Table \ref{RadiometricResults}).

\begin{figure*}[!hpbt]
   \centering
   \includegraphics[width=9.1cm]{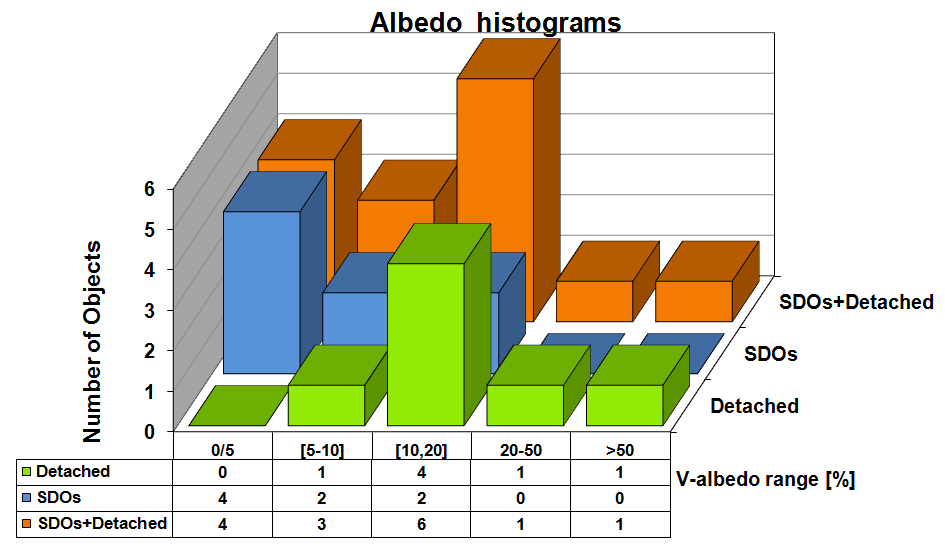}
   \includegraphics[width=9.1cm]{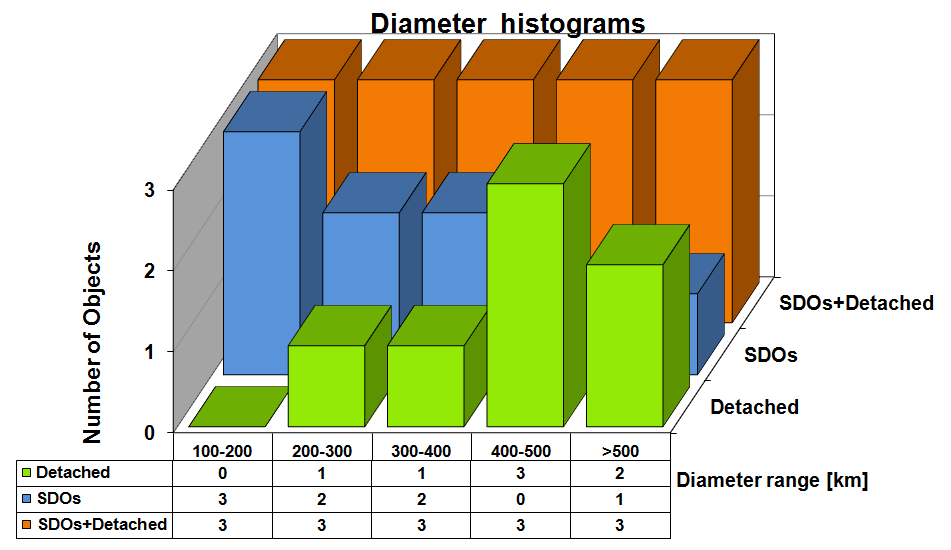}

      \caption{\textbf{Left:} geometric albedo histograms for SDOs/Detached, SDOs and detached objects. \textbf{Right:} diameter histograms for SDOs/Detached, SDOs and detached objects. 
              }
        \label{histogramas}      
   \end{figure*}

The $V$ geometric albedos for our targets vary from 3.8 \% (2002 XU$_{93}$) to 84.5 \% (Eris), with an unweighted mean value of 11.2$\pm$7.6 \% excluding Eris. If weighted by the relative errors (\textit{i.e.} $1/(\sigma/p_V)^2$), the mean albedo is 6.4$\pm$6.3 \% (for $\sigma$ we used the mean of upward and downward uncertainties on $p_V$). This error bar on the mean albedo includes the dispersion of the albedo measurements and their individual RMS errors. It is dominated by the dispersion between individual measurements, so it has a limited significance.
This variation of geometric albedos can be seen in Figure \ref{histogramas}-left. For SDOs, there is a lack of albedos$>$20~\% and a clustering  in the [0-5 \%] range. In contrast, detached objects in our sample all have albedos$>5 \%$, with a peak in the [10-20 \%] range. The unweighted/weighted mean is 6.9/5.2 \% for the SDOs and 17.0/12.3 \% for detached objects excluding Eris.  

Diameters vary from 112 km (2002 PN$_{34}$) to 2454 km (Eris), with a remarkably uniform distribution (Figure \ref{histogramas}, right). However, the 
distribution of diameters is seemingly different between SDOs and detached objects. SDO diameters are smaller than 400 km, except for an outlier 
at 1280 km (2007 OR$_{10}$), while all detached objects are larger than 200 km. This apparent tendency of detached objects
to be larger than SDOs is, however, probably an effect of the discovery bias (see discussion in Sect. \ref{observations}), as detached objects are typically discovered at larger heliocentric distances than SDOs.

Except for Eris, where the ``floating $\eta$" fit fails, the combination of \textit{Spitzer} and \textit{Herschel} data allows us to determine the beaming factor 
for 7 objects (6 SDO and 1 detached --see Table \ref{RadiometricResults}--). The (weighted by absolute errors) mean-$\eta$ obtained from these results is 1.14$\pm$0.15. This is fully consistent with the mean
1.20$\pm$0.35 $\eta$-value derived from the \textit{Spitzer} dataset (Stansberry et al. 2008) and justifies the use of this mean value for those
cases where $\eta$ cannot be determined. Beaming factors only barely larger than $\sim$1 (as opposed to $\eta >>$ 1) indicate that KBOs in general and our sample objects in particular
are close to the slow rotator regime (i.e. STM), which is a remarkable result given the very low temperatures prevailing on these objects and their usually
short rotation periods. 
Except in the unlikely situation where they are all seen pole-on, this implies very low thermal inertia and/or very
large surface roughness. Lellouch et al. (2011) determine the thermal inertia of Charon's surface to be $\Gamma$ = 10-20 J m$^{-2}$s$^{-1/2}$K$^{-1}$ (MKS)
and find this to be equivalent to $\eta$ = 1.21 - 1.40. Yet Charon is a slow rotator (6.4 day) and a rough scaling indicates that
the same $\eta$ range, applied e.g. to Haumea or Orcus, would be representative of even much smaller thermal inertias 
($\Gamma$ = 0.5-3 MKS). Hence a general conclusion is that unless they are very rough, KBO surfaces must have very low thermal conductivity,
suggesting a high degree of surface porosity. Similar thermal inertias and $\eta$'s for KBOs were obtained in M{\"u}ller et al. (2010).

\subsection{Comparison with earlier results}
Six objects of our sample (1996 TL$_{66}$, 2002 PN$_{34}$, Ceto, Typhon, 2005 TB$_{190}$ and Eris) have previously published sizes/albedos. We here briefly
compare results for the first five, and discuss Eris in more detail in next subsection. Note that 2007 UK$_{126}$ has been reported as binary system (\cite{2009DPS....41.4707N}), but there is no mass estimation for this object, for this reason so we do not discuss it here.

\textbf{(15874) 1996 TL$_{66}$}. Our preferred solution for this SDO ($D=$ 339$\pm$20 km, $p_{V}= 11.0^{+2.1}_{-1.5}$ \%,  $\eta= 1.15^{+0.08}_{-0.05}$)
is at odds with the \cite{stansberry08} results ($D= 575^{+116}_{-115}$ km, $p_{V}= 3.5^{+2.0}_{-1.0}$ \%, $\eta= 1.8\pm0.3$) based on \textit{Spitzer}-MIPS fluxes 
at 24 and 70 $\mu m$. Our reprocessing of these data leads to very different fluxes at 24 $\mu m$ (0.32 mJy vs 0.38
mJy in  \cite{stansberry08}) and especially at 70 $\mu $m (10.1 mJy vs 22 mJy). Fitting these revised \textit{Spitzer} fluxes alone gives a best fit
($D$= 328 km, $p_{V}$= 11.8 \%,  $\eta$= 1.12) in excellent agreement with our preferred solution.

\textbf{(73480) 2002 PN$_{34}$}. Although the 70 $\mu m$ fluxes from \textit{Spitzer} and \textit{Herschel} are very different (Figure \ref{MIPS_PACS_fits}), this can be
ascribed to the different observing distances. Our preferred results are: $D= 112\pm7$ km, $p_{V}= 4.9\pm0.6$ \%, $\eta= 1.02^{+0.07}_{-0.09}$,  in a good agreement with \cite{stansberry08} results: 
($D= 120\pm10$ km, $p_{V}= 4.3^{+0.8}_{-0.7}$ \%, $\eta= 1.10^{+0.16}_{-0.15}$). This object is the smallest one in our SDOs/Detached sample.

\textbf{(65489) Ceto/Phorcys} is a binary SDO, previously observed with \textit{Spitzer}-MIPS (\cite{2007Icar..191..286G}, \cite{stansberry08}). 
Model fits indicate that the optimum model ($D= 281\pm11$ km, $p_{V}= 5.6\pm0.6$ \%, $\eta= 1.04\pm0.05$) matches the 24 $\mu$m flux but somewhat overestimates the \textit{Spitzer} 70 $\mu$m flux, indicative of a colder object than the \textit{Spitzer} data would suggest. 
It is therefore not surprising that the \textit{Spitzer}-only solution of \cite{stansberry08} has a 
lower $\eta$ ($0.86^{+0.10}_{-0.09}$), lower $D$ ($230^{+19}_{-18}$ km) and higher $p_{V}$ (7.7$^{+1.4}_{-1.1}$ \%).

Grundy et al. (2007) estimate a mass of (5.41$\pm$0.42)$\times$10$^{18}$ kg for the Ceto/Phorcys system. Using an equivalent diameter of 218$^{+20}_{-22}$ km, they obtain a bulk density of 1.37$^{+0.66}_{-0.32}$ g cm$^{-3}$. We derive a smaller bulk density of  0.64$^{+0.16}_{-0.13}$ g cm$^{-3}$ due to our larger equivalent diameter ($D= 281$ km which gives $D_{Ceto}=$ 223$\pm$10 km, and $D_{Phorcys}=$ 171$\pm$10 km, assuming the same albedo for the two components and $\Delta m=$ 0.58$\pm$0.03 mag). 
This low bulk density is compatible with a porous surface, which is also supported by the very low $\eta= 1.04$ value.
 
\textbf{(42355) Typhon/Echidna} is a binary SDO. Our preferred results are $D= 185\pm7$ km, $p_{V}= 4.4\pm0.3 \%$, $\eta= 1.48\pm0.07$, consistent with the Stansberry et al. (2008) solution obtained by applying a fixed $\eta$ = 1.2$\pm$0.35 model to the 24 $\mu m$ flux 
($D= 174^{+16}_{-18}$ km, $p_{V}= 5.1^{+1.2}_{-0.8} \%$). In contrast, these results are inconsistent with the preliminary
\textit{Herschel} results ($D= 138\pm9$ km and $p_{V}= 8.0 \pm 1.0$ \%), based on SDP data (\cite{2010A&A...518L.146M}).
The latter observations were performed in chop-nod mode (less sensitive than scan-map for point sources) and led to lower 70 and 100 $\mu m$ fluxes
and no detection at 160 $\mu m$.

Grundy et al. (2008) obtain a mass of (9.49$\pm$0.52)$\times$10$^{17}$ kg for this binary system. Using an equivalent diameter of 179$^{+16}_{-18}$ km, they obtain a bulk density of 0.44$^{+0.44}_{-0.17}$ g cm$^{-3}$. We derive a nominally even smaller (but compatible within error bars) bulk density of 0.36$^{+0.08}_{-0.07}$ g cm$^{-3}$ from a slightly larger equivalent diameter ($D= 185$ km, \textit{i.e.} $D_{Typhon}=$ 162 $\pm$7 km, and $D_{Echidna}=$ 89 $\pm$ 6 km, assuming same albedo 
and $\Delta m=$ 1.30$\pm$0.06 mag). Again, this extremely low bulk density points to a porous body. 

\textbf{(145480) 2005 TB$_{190}$}. This detached object was not observed by \textit{Spitzer}-MIPS, but was already measured in chop-nod mode by
\textit{Herschel}. Our new results, using a fixed-$\eta$ mode  ($D= 464 \pm 62$ km, $p_{V}= 14.8^{+5.1}_{-3.6} \%$) are reasonably consistent (to within 1-$\sigma$)
with the preliminary values published by M{\"u}ller et al. (2010) (D= $375 \pm 45$ km, $p_{V}= 19.0 \pm 5.0 \%$).


\subsection{Two large objects, 2007 OR$_{10}$ and Eris}
\label{large_objects}

\textbf{2007 OR$_{10}$}. Its $H_R$ magnitude of $\sim$2 indicates that this very distant object (currently at 86.3 AU) must be one of the 
largest TNOs, unless it exhibits a very high albedo. Brown et al. (2011)
recently reported that its spectrum is markedly red in the near-infrared and shows features due to water ice. This is an unsual combination as
most KBOs with water ice absorptions have neutral colors, but this characteristic is shared by Quaoar. Although their data are unsufficient to show a detection of methane ice, 
Brown et al. (2011) used the analogy with Quaoar, where methane ice is present (\cite{2007ApJ...670L..49S}), to suggest that the red color of 2007 OR$_{10}$ 
is due to irradiation of small amounts of methane ice. They further assumed an albedo equal to that of Quaoar (nominally 0.18 for an adopted Quaoar radius
of 960 km, but allowing for 50 \% uncertainties), and showed, based on the Schaller and Brown (2007b) sublimation model, that  2007 OR$_{10}$ is in a similar 
volatile retention regime as Quaoar, with the main volatile species (N$_2$, CO, CH$_4$) being marginally stable over the solar system age.

2007 OR$_{10}$ is clearly detected in the three PACS channels. In the lack of $V$ measurements, we assumed $H_{R}$ = 1.96 $\pm$ 0.16 magnitudes based on the fit of eight good quality R-measurements from MPC data (see Table \ref{tableOBS}). A fixed-$\eta$ fit is satisfactory (Figure \ref{PACS_fits}), providing $D= 1280 \pm 210$ km, $p_{R}= 18.5^{+7.6}_{-5.2} \%$.
This puts the object in the Charon, Makemake, Haumea, etc... size class, i.e. a plausible candidate for dwarf planet status. To convert $p_R$ to $p_V$ we use equation (\ref{albedoconversion}) from Sect. \ref{Hmag_issue}, but use Quaoar's V-R value = 0.67$\pm$0.02 (\cite{2004A&A...421..353F}). This yields $p_{V}= 13.9^{+5.7}_{-3.9} \%$,
nicely consistent with the value assumed by Brown et al. (2011). 2007 OR$_{10}$ thus indeed appears as a sibling to Quaoar, probably of even larger size.

\textbf{(136199) Eris.} Although Eris was observed by \textit{Spitzer}, it was not detected at 24 $\mu$m (with a 0.005 mJy upper limit), preventing us
from determining its beaming factor. Assuming therefore $\eta$=1.20 $\pm$ 0.35, we obtain a best fit solution for  $D= 2454\pm117$ km, $p_{V}= 84.5 \pm 8.8$ \%. Using the mass determined by Brown et al. (2007) --recomputed in Grundy et al. (2011)-- for the Eris/Dysnomia system  ((1.688$\pm$0.035)$\times$10$^{22}$ kg ) we obtain a bulk density of 2.23$^{+0.41}_{-0.33}$ g cm$^{-3}$, assuming the same albedo for Eris and Dysnomia ($D_{Eris}$= 2434 $\pm$ 117 km, and $D_{Dysnomia}$= 316 $\pm$ 23 km). 
It is probably more realistic to assume that Dysnomia's surface is much darker than Eris'. Assuming for definiteness a 5 times lower albedo, we obtain $D_{Eris}$=2356 $\pm$ 117 km and $D_{Dysnomia}$= 685 $\pm$ 50 km, yielding a global density of 2.40$^{+0.46}_{-0.37}$ g cm$^{-3}$,
similar to the Brown et al. (2007) value (2.3 $\pm$ 0.3 g cm$^{-3}$).

Near-IR spectroscopy (\cite{2010ApJ...725.1296T}, \cite{2009AJ....137..315M}, \cite{2005ApJ...635L..97B}) indicates
that Eris' surface composition is very similar to Pluto's, with the presence of methane ice along with another dominant ice, presumably
nitrogen, suggestive of a high albedo. Prior to our measurements, Eris' size and albedo have been determined by several groups and methods. 
The first measurement was achieved by \cite{Bertoldi2006}  who used IRAM thermal observations at 1200 $\mu m$  and obtained $D= 3000^{+300}_{-100}$ km
and $p_{V}= 60 \pm 10$ \%. Subsequently, \cite{Brown2006} estimated $D= 2400 \pm 100$ km, $p_{V}= 86 \pm 7$ \% from direct imaging with the Hubble Space Telescope. Based on the
\textit{Spitzer} 70 $\mu m$ flux, \cite{stansberry08} obtained $D= 2657^{+216}_{-209}$ km, $p_{V}= 68.9^{+12.2}_{-10.0}
 \%$ from \textit{Spitzer}-MIPS and hybrid-STM fit. 
Within uncertainties, our results agree with all these values except with those of \cite{Bertoldi2006}, which overestimate the size and underestimate the albedo. 
Most recently, a stellar occultation by Eris on November 6th 2010 
provided a very accurate determination of Eris' diameter and albedo: $D= 2326 \pm 12$ km, $p_{V}= 96^{+9}_{-4}$ \% (\cite{2011Natur.478..493S}), assuming
a spherical shape. (Note however that allowing for elliptical shape, Sicardy et al. find a broader range of solutions,  68.3 \% (1-$\sigma$) of which
have an effective diameter of 2330 $\pm$ 180 km). 
Eris' diameter is thus indistinguishable from Pluto's given error bars on both, supporting the idea that Eris is a Pluto twin,
with an even brighter surface, presumably due to a collapsed atmosphere at aphelion. 

At face value and given error bars, our result on Eris' diameter agrees nicely with that of Sicardy et al. (2011), especially if we assume that Dysnomia is 
intrinsically much darker than Eris. Similarly, our density for the system is in line with theirs (2.52 $\pm$ 0.05 g cm$^{-3}$). However, the consistency is 
actually not complete.
Sicardy et al. (2011) reanalyzed the \textit{Spitzer} 70 $\mu m$ and IRAM 1200 $\mu m$ fluxes in the light of their determined diameter and albedo. They considered both a standard thermal model (STM) and an isothermal latitude model (ILM), introducing an ``$\eta$"(= $\eta_{R}$) parameter describing exclusively surface roughness effects (and ranging from $\eta_{R}=$ 1.0 --no roughness-- to $\eta_{R}=$ 0.7 --large roughness--). In this framework, the only free parameter of their thermal model was the phase integral $q$. They found that the thermal measurements implied $q$ = 0.49-0.66 for the STM case (fully consistent with the value of Saturn's brightest satellites), but an implausible $q$ = 0-0.24 for the ILM. On this basis, the STM was strongly favored, implying either a pole-on orientation or a negligible thermal inertia. The inconsistency with our analysis resides in the fact that we assumed $\eta$= 1.20 $\pm$ 0.35 and that our $p_{V}$ solution implies $q$ = 0.75, according to the assumed  Brucker et al. (2009) expression. Another way of seeing this is to note that if we now assume the two limiting STM cases considered by \cite{2011Natur.478..493S} (i.e. q= 0.49, $\eta_{R}=$ 1.0, and q= 0.66, $\eta_{R}=$ 0.7) and re-fit the PACS + MIPS data, we obtain best fit diameters of 2034 km and 2139 km respectively, and V-albedos larger than 100\% in the two cases, inconsistent with the occultation results. Ultimately, the discrepancy can be traced to the fact that Sicardy
et al. (2011) included the IRAM 1200 $\mu$m flux in their model, which as they noted tends to indicate warmer temperatures than the 
\textit{Spitzer} 70 $\mu m$ flux, while on the contrary our best fit model tends to underestimate the latter. The resolution of this problem will 
require a simultaneous fits of all datasets, considering in particular (i) non-spherical shapes (ii) separate thermal models for Eris and Dysnomia and (iii) more elaborate thermophysical models in which the thermal inertia, surface roughness and spin orientation are free parameters. This study is deferred to a subsequent paper (Kiss et al. \textit{in prep.}-b). Note finally that the interpretation of thermal observations of a volatile-covered body such as Eris is expected to be even more complex 
than presented here. An additional complication is that volatile transport will result in surface variegation, i.e. multiple
surface albedos. This is demonstrated in the case of Pluto, which does show large albedo spatial variations and whose thermal
data require three-terrain model (\cite{2011Icar..214..701L}, \cite{Mommert2011}). Another example is Makemake, the visual appearance 
of which is unknown but for which a two-terrain model is required to fit the radiometric data (\cite{Lim2010}).

\begin{table*}

\caption{Radiometric fit results for the objects sample. These are the results obtained by the application of an hybrid-STM to the fluxes obtained by \textit{Herschel}-PACS (with \textit{Spitzer}-MIPS data when are available). Preferred results for each target are in bold} 

\label{RadiometricResults} 

\centering 

\begin{tabular}{r c c c c c} 

\hline\hline 

Object	&	Data	&	$D$ [km]	&	$p_{v}$ [$\%$]	&	$\eta$	&	Class.	\\


\hline 

(15874) 1996 TL$_{66}$	&	PACS	&	341$\pm$40	&	11.0$^{+3.8}_{-2.5}$	&	1.20$\pm$0.35*	&	SDO	\\%
\textbf{(15874) 1996 TL$_{66}$}	&	\textbf{PACS/MIPS}	&	\textbf{339$\pm$20}	&	\textbf{11.0$^{+2.1}_{-1.5}$}	&	\textbf{1.15$^{+0.08}_{-0.05}$}	&	\textbf{SDO}\\%
(82158) 2001 FP$_{185}$	&	PACS	&	329$\pm$42	&	4.7$^{+1.3}_{-1.0}$	&	1.20$\pm$0.35*	&	SDO	\\%
\textbf{(82158) 2001 FP$_{185}$}	&	\textbf{PACS/MIPS}	&	\textbf{332$^{+31}_{-24}$}	&	\textbf{4.6$\pm$0.7}	&	\textbf{1.23$^{+0.24}_{-0.19}$}	&	\textbf{SDO}	\\%
(73480) 2002 PN$_{34}$	&	PACS	&	110$\pm$12	&	5.2$^{+1.3}_{-0.9}$	&	1.20$\pm$0.35*	&	SDO	\\%
\textbf{(73480) 2002 PN$_{34}$}	&	\textbf{PACS/MIPS}	&	\textbf{112$\pm$7}	&	\textbf{4.9$\pm$0.6}	&	\textbf{1.02$^{+0.07}_{-0.09}$}	&	\textbf{SDO}	\\%
(127546) 2002 XU$_{93}$	&	PACS	&	173$\pm$20	&	3.5$^{+0.9}_{-0.6}$	&	1.20$\pm$0.35*	&	SDO	\\%
\textbf{(127546) 2002 XU$_{93}$}	&	\textbf{PACS/MIPS}	&	\textbf{164$\pm$9}	&	\textbf{3.8$\pm$0.4}	&	\textbf{1.12$^{+0.05}_{-0.08}$}	&	\textbf{SDO}	\\%
\textbf{2007 OR$_{10}$}	&	\textbf{PACS}	&	\textbf{1280$\pm$210}  &	\textbf{18.5$^{+7.6}_{-5.2}\star$}	&	\textbf{1.20$\pm$0.35*}	&	\textbf{SDO}	\\%
\textbf{2007 RW$_{10}$}	&	\textbf{PACS}	&	\textbf{247$\pm$30}	&	\textbf{8.3$^{+6.8}_{-3.9}\star$}	&	\textbf{1.20$\pm$0.35*}	&	\textbf{SDO}	\\%
(65489) Ceto	&	PACS	&	300$\pm$39	&	4.9$^{+1.6}_{-1.0}$	&	1.20$\pm$0.35*	&	SDO	\\%
\textbf{(65489) Ceto}	&	\textbf{PACS/MIPS}	&	\textbf{281$\pm$11}	&	\textbf{5.6$\pm$0.6}	&	\textbf{1.04$\pm$0.05}	&	\textbf{SDO}	\\%
(42355) Typhon	&	PACS	&	175$\pm$17	&	4.9$^{+1.1}_{-0.8}$	&	1.20$\pm$0.35*	&	SDO	\\%
\textbf{(42355) Typhon}	&	\textbf{PACS/MIPS}	&	\textbf{185$\pm$7}	&	\textbf{4.4$\pm$0.3}	&	\textbf{1.48$\pm$0.07}	&	\textbf{SDO}	\\%
\textbf{(40314) 1999 KR$_{16}$}	&	\textbf{PACS}	&	\textbf{254$\pm$37}	&	\textbf{20.4$^{+7.0}_{-5.0}\star$}	&	\textbf{1.20$\pm$0.35*}	&	\textbf{DO}	\\%
(120132) 2003 FY$_{128}$	&	PACS	&	479$^{+55}_{-67}$	&	7.2$^{+2.7}_{-1.5}$	&	1.20$\pm$0.35*	&	DO	\\%
\textbf{(120132) 2003 FY$_{128}$}	&	\textbf{PACS/MIPS}	&	\textbf{460$\pm$21}	&	\textbf{7.9$\pm$1.0}	&	\textbf{1.07$\pm$0.08}	&	\textbf{DO}	\\%
\textbf{2005 QU$_{182}$}	&	\textbf{PACS}	&	\textbf{416$\pm$73}	&	\textbf{32.8$^{+16.0}_{-10.9}$}	&	\textbf{1.20$\pm$0.35*}	&	\textbf{DO}	\\%
\textbf{(145480) 2005 TB$_{190}$}	&	\textbf{PACS}	&	\textbf{464$\pm$62}	&	\textbf{14.8$^{+5.1}_{-3.6}$}	&	\textbf{1.20$\pm$0.35*}	&	\textbf{DO}	\\%
\textbf{2007 OC$_{10}$}	&	\textbf{PACS}	&	\textbf{309$\pm$37}	&	\textbf{12.7$^{+4.0}_{-2.8}$}	&	\textbf{1.20$\pm$0.35*}	&	\textbf{DO}	\\%
\textbf{(229762) 2007 UK$_{126}$}	&	\textbf{PACS}	&	\textbf{599$\pm$77}	&	\textbf{16.7$^{+5.8}_{-3.8}$}	&	\textbf{1.20$\pm$0.35*}	&	\textbf{DO}	\\%
(136199) Eris	&	PACS	&	2420$^{+100}_{-119}$	&	86.7$^{+9.8}_{-6.3}$	&	1.20$\pm$0.35*	&	DO	\\%
\textbf{(136199) Eris}	&	\textbf{PACS/MIPS}	&	\textbf{2454$\pm$117}	&	\textbf{84.5$\pm$8.8}	&	\textbf{1.20$\pm$0.35*}	&	\textbf{DO}	\\%

\hline 

\end{tabular}

\footnotesize{\textbf{Object:} SDO or detached object observed with \textit{Herschel}-PACS (and \textit{Spitzer}-MIPS when available). \textbf{Data:} PACS, model results using only fluxes from \textit{Herschel}-PACS. PACS/MIPS, model results using fluxes from \textit{Herschel}-PACS and \textit{Spitzer}-MIPS. \textbf{$D$ [km]:} Diameter and uncertainty, in kilometres, obtained from the thermal model. \textbf{$p_{v}$ [$\%$]:} geometric albedo and uncertainty, in V-band, obtained from the thermal model. $\star$ indicates geometric albedos derived from H$_{R}$ magnitudes instead of H$_{V}$, (i.e, p$_{r}$ [$\%$]). \textbf{$\eta$:} Parameter from the thermal model application. * objects for which there are not \textit{Spitzer}-MIPS fluxes available, in this case we assume a mean $\eta$ value to perform the radiometric fits. \textbf{Class.}, dynamical classification following Gladman et al. 2008.  SDO = scattered disc object, DO = detached object.}

\end{table*}
	

\subsection{Search for correlations}

\subsubsection{Description of the method}

Using Spearman-rank correlation coefficient (\cite{Spearman1904}), also
known as Spearman-$\rho$, we have examined our results for correlations
between albedo ($p_v$), beaming factor ($\eta$), diameter ($D$), absolute
magnitude ($H$), color ($B-R$), perihelion distance ($q$), orbital
inclination ($i$),
semi-major axis ($a$), and heliocentric distance at the time of
observation ($r_h$).

Spearman-$\rho$ is distribution-free and less sensitive to outliers than
most other methods but, like others, treats data-points as `exact' and
does not take into account their possible error bars. Any variations in
the measured data points within their error bars, may change the
correlation coefficient, however. Furthermore, each correlation
coefficient has its own confidence interval, which depends on the number
of data points and on the magnitude of the correlation. We use procedures described in Appendix B (Sect. \ref{spearmanconerrors}) 
and in Peixinho et al (2004) to deal with these effects to obtain Spearman-$\rho$ values that take into account the error bars.

We call correlations ``strong" when $|\rho|\geqslant$ 0.6.
Efron \& Tibshirani (1993) analyze the evidence for correlation with the follow criteria (where $CL$ is the confidence level defined in Appendix B): $CL\geqslant99\%$ ($p\leqslant$ 0.010):
very strong evidence of correlation; and $CL\geqslant99.7\%$ ($p\leqslant$ 0.003), the `standard' $3\sigma$ significance: clear correlation. We restrict ourselves to the cases with $CL\geqslant99.5\%$ ($p\leqslant$ 0.005), i.e. at least 2.8$\sigma$ confidence.

\begin{figure*}[!hpbt]
   \centering
   \includegraphics[width=4.48cm,angle=90]{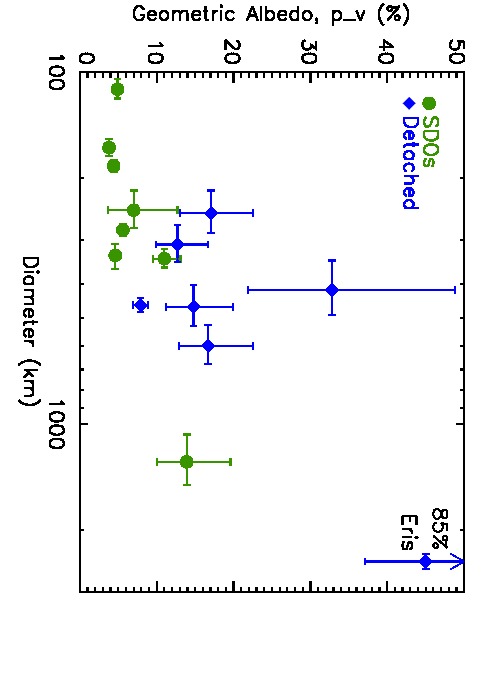}
   \includegraphics[width=4.48cm,angle=90]{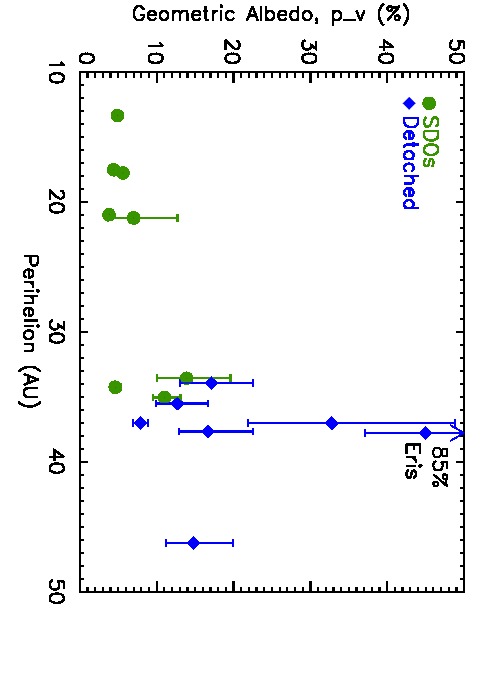}
   \includegraphics[width=4.48cm,angle=90]{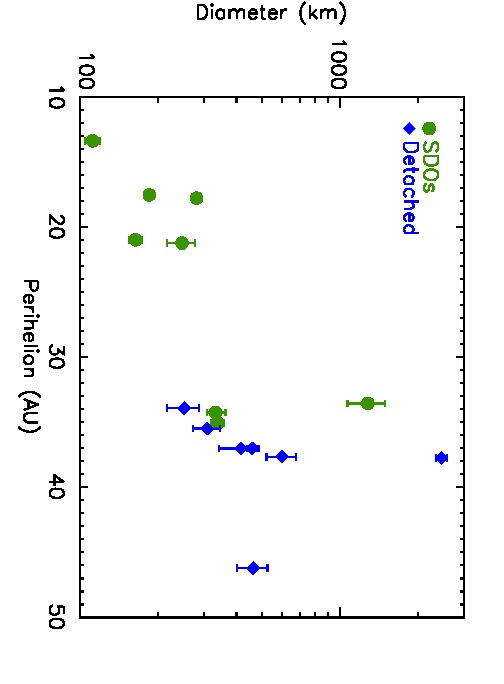}
      \caption{Plots illustrating the three more interesting correlations with $CL\geqslant99.5\%$ ($\geqslant2.8\sigma$) in our results. 
      From left to right:
albedo \textit{vs} diameter, albedo \textit{vs} perihelion distance,
and diameter \textit{vs} perihelion distance (Eris is out of the
albedo scale in the first two panels, but its albedo is indicated).}
         \label{correlationplots}
   \end{figure*}
   
\subsubsection{Correlations analysis}

Based on the previously described methods, the strongest and/or more interesting correlations are shown in Table \ref{correlations}. These are
given for the whole sample (including or excluding Eris), and for SDOs and detached objects separately. Before discussing them, we stress that results must be taken with care because of small number statistics. In general, more significant correlations are found for SDOs than for detached objects.

``Clear correlations'' ($\geqslant99.7\%$) in Table \ref{correlations} are seen between $D$ and $H$ (negative correlation, 3.0$\sigma$), $p_V$ and $H$ (negative, 4.2$\sigma$), $D$ and $r_h$ (positive, 4.9$\sigma$), $p_V$ and $r_h$ (positive, 3.8$\sigma$). The first two anticorrelations are a simple consequence of equation (1), with no physical implication. 
The positive correlations between $D$ and $r_h$, and between $p_V$ and $r_h$, are most probably biases due to object selection criterion and 
discovery bias, as discussed in Sect. \ref{observations}.

Additional ``clear'' or ``very strong evidence''  ($\geqslant99\%$) for positive correlations are found between $D$ and $q$ (3.4$\sigma$) and $p_v$ and
$q$ (2.9$\sigma$) (Figure \ref{correlationplots}). There is no a priori relationship between diameter, albedo and perihelion distance, 
hence these correlations could plausibly be real, as discussed below. However, they could also reflect biases due to the
correlation between $q$ and $r_h$ (at 3.9$\sigma$ for our sample, see Table \ref{correlations}), i.e. the discovery
bias discussed in Sect. \ref{observations}. 
On the other hand, it is noteworthy that no high albedo objects (e.g. $p_v$ $>$10 \%) are seen at small heliocentric or perihelion distance ($\leq$30 AU), which should not be a bias. Note that when SDOs and detached objects are treated separately, the  $D$ vs $q$ correlation remains significant at 2$\sigma$ only, while the $p_v$ and $q$ correlation essentially vanishes. Correlations also decrease (marginally) if Eris is excluded from the sample.

Based on {\em Spitzer} measurements of about 50 objects, Stansberry et al. (2008) noted the apparent positive correlation between $p_v$ and $q$
(3.5$\sigma$ in their case). While recognizing the above discovery bias, they proposed it to be real, offering two explanations. First, objects near the Sun tend to experience higher temperatures, hence may be subject to increased sublimation of their intrinsically bright ices. Conversely, increased UV-photolysis and solar wind radiolysis associated with the low heliocentric/perihelion distances may preferentially darken the surfaces of these objects. It seems more difficult to find a physical explanation for an hypothetical correlation between diameter and perihelion distance, especially because the current orbits of the SDOs/Detached objects tell us little if anything about the formation region. This apparent correlation is most probably a bias, as discussed above. Note also that unlike Stansberry et al. (2008) we do not find any (even tentative) correlation between $p_v$ and the semi-major axis $a$ ($\rho$= 0.21, 0.8$\sigma$, see Table \ref{correlations}).

We find a very strong correlation between albedo and diameter ($p_{V}$ \textit{vs.} $D$) when SDOs are included, in particular for the whole sample at 2.9$\sigma$ significance (see Figure \ref{correlationplots}). This result was obtained by Stansberry et al. (2008) for KBOs (defined there as any object but Centaurs) but was not robust against small changes in the classification. This correlation implies a trend for larger objects having higher albedos. An explanation could invoke the fact that larger objects can retain bright ices (released either by collision or intrinsic activity) more easily than small objects do.
Eris follows this tendency, although its extremely high albedo puts it as an outlier in the plot of Figure \ref{correlationplots}.
An important aspect here is that there are no small objects ($\leq$300 km) with high albedo ($\geq$10 \%). This cannot be a bias, as such objects would have been targetted in our sample, yet poorly detected. Note that if objects with albedos $\geq$10 \% are to be associated with the presence of surface ices, then such ices should be only moderately volatile (i.e. CH$_4$ or less), as volatile retention models (\cite{2007ApJ...659L..61S}, \cite{2011ApJ...738L..26B}) show that only the largest ($\geq$1000-1500 km) bodies can retain the most volatile ices such as N$_2$ or CO over the solar system age.

Finally, we do not detect any correlation between diameter/albedo and color, or inclination, nor between the beaming factor $\eta$ with any parameter (though the sample is limited to 7 objects in the latter case).

In summary, in spite of the relatively small sample covered in this paper, we find three possible correlations (\textit{i.e.} $p_{V}$ \textit{vs.} $D$, $p_{V}$ \textit{vs.} $q$, and $D$ \textit{vs.} $q$), of which at least the first one --large objects tend to have brighter surfaces than small objects-- is real and may be physically explained. More subtle correlations may be found, but with lower significance, and larger datasets will be required for confirmation.

\begin{table*}[!hpbt]
\begin{center}
\caption{Summary of correlations with diameter and geometric albedo}
\label{correlations}
\renewcommand{\arraystretch}{1.2}
\setlength{\tabcolsep}{0.75mm}
\begin{tabular}[]{llrrcrlrl}
\hline \hline
                  &            &  & \multicolumn{1}{c}{Without}    & & \multicolumn{4}{c}{Accounting for} \\
Variables & Class & & \multicolumn{1}{c}{error bars} & & \multicolumn{4}{c}{data error bars} \\
\cline{4-4} \cline{6-9}
 &  & n & \multicolumn{1}{c}{$\rho$} & & \multicolumn{1}{c}{$\langle \rho \rangle^{+\sigma}_{-\sigma}$} & \multicolumn{1}{c}{$p$} & $CL(\%)$ & \multicolumn{1}{c}{($\sigma_p$)}\\
\hline
${ D }$ {\it vs.} ${ H }$ & SDOs+DOs      & $  15 $ & $ -0.93 $ & & $ -0.74 ^{+ 0.48 }_{- 0.19 }$ & $ 0.00244  $ &  99.8    & $( 3.03 )$ \\
						  & SDOs+DOs*     & $  14 $ & $ -0.91 $ & & $ -0.68 ^{+ 0.37 }_{- 0.19 }$ & $ 0.00989  $ &  99.0    & $( 2.58 )$ \\
						  & SDOs          & $   8 $ & $ -0.93 $ & & $ -0.80 ^{+ 0.38 }_{- 0.14 }$ & $ 0.0281   $ &  97.2    & $( 2.20 )$ \\
 						  & DOs           & $   7 $ & $ -0.86 $ & & $ -0.53 ^{+ 0.49 }_{- 0.29 }$ & $ 0.243    $ &  75.7    & $( 1.17 )$ \\
\hline    
${ D }$ {\it vs.} ${r_h}$ & SDOs+DOs      & $  15 $ & $  0.93 $ & & $  0.92 ^{+ 0.05 }_{- 0.11 }$ & $ 0.000001 $ &  $>$99.9 & $( 4.91 )$ \\
						  & SDOs+DOs*     & $  14 $ & $  0.92 $ & & $  0.90 ^{+ 0.06 }_{- 0.14 }$ & $ 0.000010 $ &  $>$99.9 & $( 4.42 )$ \\
						  & SDOs     	  & $   8 $ & $  0.98 $ & & $  0.96 ^{+ 0.03 }_{- 0.09 }$ & $ 0.00134  $ &  99.9    & $( 3.21 )$ \\		  
						  & DOs           & $   7 $ & $  0.71 $ & & $  0.74 ^{+ 0.25 }_{- 1.26 }$ & $ 0.0782   $ &  92.2    & $( 1.76 )$ \\
 						  & DOs*          & $   6 $ & $  0.54 $ & & $  0.58 ^{+ 0.26 }_{- 0.49 }$ & $ 0.256    $ &  74.4    & $( 1.14 )$ \\
\hline    
${ D }$ {\it vs.} ${ q }$ & SDOs+DOs      & $  15 $ & $  0.81 $ & & $  0.79 ^{+ 0.11 }_{- 0.22 }$ & $ 0.000594 $ &  $>$99.9 & $( 3.43 )$ \\
						  & SDOs+DOs*     & $  14 $ & $  0.78 $ & & $  0.76 ^{+ 0.14 }_{- 0.28 }$ & $ 0.00270  $ &  99.7    & $( 3.00 )$ \\
						  & SDOs          & $   8 $ & $  0.81 $ & & $  0.80 ^{+ 0.11 }_{- 0.23 }$ & $ 0.0264   $ &  97.4    & $( 2.22 )$ \\
						  & DOs           & $   7 $ & $  0.86 $ & & $  0.80 ^{+ 0.20 }_{- 1.39 }$ & $ 0.0512   $ &  94.9    & $( 1.95 )$ \\
						  & DOs*          & $   6 $ & $  0.89 $ & & $  0.80 ^{+ 0.13 }_{- 0.33 }$ & $ 0.106    $ &  89.4    & $( 1.61 )$ \\
\hline						      
\hline	
${ p_V }$ {\it vs.} ${ H }$ & SDOs+DOs    & $  15 $ & $ -0.88 $ & & $ -0.87 ^{+ 0.11 }_{- 0.06 }$ & $ 0.000027 $ &  $>$99.9 & $( 4.20 )$ \\		      
							& SDOs+DOs*   & $  14 $ & $ -0.85 $ & & $ -0.84 ^{+ 0.13 }_{- 0.07 }$ & $ 0.000186 $ &  $>$99.9 & $( 3.74 )$ \\									    & SDOs        & $   8 $ & $ -0.81 $ & & $ -0.77 ^{+ 0.34 }_{- 0.15 }$ & $ 0.036888 $ &  96.3    & $( 2.09 )$ \\									    & DOs         & $   7 $ & $ -0.64 $ & & $ -0.66 ^{+ 0.45 }_{- 0.22 }$ & $ 0.128142 $ &  87.2    & $( 1.52 )$ \\
\hline    
${ p_V }$ {\it vs.} ${ D }$ & SDOs+DOs    & $  15 $ & $  0.71 $ & & $  0.71 ^{+ 0.14 }_{- 0.22 }$ & $ 0.004045 $ &  99.6    & $( 2.88 )$ \\
						    & SDOs+DOs*   & $  14 $ & $  0.65 $ & & $  0.64 ^{+ 0.16 }_{- 0.24 }$ & $ 0.015648 $ &  98.4    & $( 2.42 )$ \\
						    & SDOs        & $   8 $ & $  0.71 $ & & $  0.68 ^{+ 0.20 }_{- 0.42 }$ & $ 0.078503 $ &  92.1    & $( 1.76 )$ \\
							& DOs         & $   7 $ & $  0.25 $ & & $  0.32 ^{+ 0.57 }_{- 0.96 }$ & $ 0.483703 $ &  51.6    & $( 0.70 )$ \\
\hline    
${ p_V }$ {\it vs.} ${r_h}$ & SDOs+DOs    & $  15 $ & $  0.87 $ & & $  0.83 ^{+ 0.08 }_{- 0.14 }$ & $ 0.000126 $ &  $>$99.9 & $( 3.83 )$ \\
						    & SDOs+DOs*   & $  14 $ & $  0.84 $ & & $  0.79 ^{+ 0.10 }_{- 0.16 }$ & $ 0.001327 $ &  99.9    & $( 3.21 )$ \\
						    & SDOs        & $   8 $ & $  0.69 $ & & $  0.65 ^{+ 0.23 }_{- 0.48 }$ & $ 0.098781 $ &  90.1    & $( 1.65 )$ \\
							& DOs         & $   7 $ & $  0.68 $ & & $  0.69 ^{+ 0.28 }_{- 1.19 }$ & $ 0.105315 $ &  89.5	& $( 1.62 )$ \\
\hline    
${ p_V }$ {\it vs.} ${ q }$ & SDOs+DOs    & $  15 $ & $  0.75 $ & & $  0.71 ^{+ 0.12 }_{- 0.19 }$ & $ 0.003965 $ &  99.6    & $( 2.88 )$ \\
						    & SDOs+DOs*   & $  14 $ & $  0.71 $ & & $  0.68 ^{+ 0.15 }_{- 0.23 }$ & $ 0.010510 $ &  98.9    & $( 2.56 )$ \\
						    & SDOs        & $   8 $ & $  0.48 $ & & $  0.47 ^{+ 0.31 }_{- 0.49 }$ & $ 0.251698 $ &  74.8    & $( 1.15 )$ \\
						    & DOs         & $   7 $ & $  0.25 $ & & $  0.36 ^{+ 0.45 }_{- 0.72 }$ & $ 0.430091 $ &  57.0    & $( 0.79 )$ \\
\hline
${ p_V }$ {\it vs.} ${ a }$   & SDOs+DOs  & $  15 $ & $  0.24 $ & & $  0.21 ^{+ 0.28 }_{- 0.32 }$ & $ 0.449038 $ &  55.1    & $( 0.76 )$ \\

\hline
${ p_V }$ {\it vs.} ${ e }$ & SDOs+DOs & $ 15 $  & $ -0.41 $ & & $ -0.39 ^{+ 0.28 }_{- 0.22 }$ & $ 0.154918 $ & 84.5  & $( 1.42 )$ \\
\hline
${ p_V }$ {\it vs.} ${ i }$ & SDOs+DOs & $ 15 $  & $ 0.05 $  & & $  0.09 ^{+ 0.36 }_{- 0.38 }$ & $ 0.769100 $ & 23.1  & $( 0.29 )$ \\

\hline
\hline
${ r_h }$ {\it vs.} ${ q }$   & SDOs+DOs  & $  15 $ & $  0.81 $ & & $  0.84 ^{+ 0.09 }_{- 0.20 }$ & $ 0.000084 $ &  $>$99.9 & $( 3.93 )$ \\
\hline
\end{tabular}
\end{center}
\footnotesize{Spearman-rank correlation tests given the main correlations between the various magnitudes. *Indicates that Eris is excluded. 
\textbf{n:} number of data points. 
\textbf{$\rho$:} Spearman-rank correlation without data error bars. 
\textbf{$\langle \rho \rangle^{+\sigma}_{-\sigma}$:} Spearman-rank correlation accounting for data error bars. 
\textbf{$p$:} significance p-value of the correlation. 
\textbf{$CL(\%)$:} confidence level defined as (1-$p$)$\cdot$100\%. 
\textbf{$\sigma_p$:} Gaussian-$\sigma$ equivalent to the significance p-value (see text for details). 
\textbf{$D$:} diameter. 
\textbf{$r_h$:} heliocentric distance at the time of observation. 
\textbf{$a$:} orbital semi-major axis. 
\textbf{$e$:} orbital eccentricity. 
\textbf{$i$:} orbital inclination with respect to the ecliptic plane. 
\textbf{$q$:} perihelion. 
\textbf{$p_V$:} V-band geometric albedo. }
The last line shows that $r_h$ and $q$ are strongly correlated in our sample.
\end{table*}


\section{Summary}

We have determined albedos/sizes of 8 SDOs and 7 detached objects using \textit{Herschel} Space Observatory-PACS fluxes
 measurements at 70, 100 and 160 $\mu m$  and \textit{Spitzer}-MIPS fluxes at 23.68 and 71.42 $\mu m$  when available.
The main results, which include nine first determinations,  are summarized hereafter:

   \begin{itemize}
        
      \item{Diameters range from 100 to 2400 km. The overall diameter distribution in our sample is remarkably uniform, but
apparently different for SDOs vs detached objects. This difference might however just reflect a sample/discovery bias, 
as our sample favors large objects at large heliocentric distances, and detached objects are typically discovered 
at larger distances than SDOs.} 

	  \item{V-geometric albedos range from 3.8 to 84.5 \%. Unweighted mean albedo for the whole sample is 11.2 \%  and 6.9/17.0 \% for the SDOs/Detached (excluding Eris) respectively. There is a lack of albedos $>$ 20\% with a peak in the [0-5\%] range for SDOs. Detached objects have albedos $>$ 5 \%, with a peak in the [10-20 \%] range.}
      
      \item{We re-determine bulk densities of three binary systems, based on our estimated sizes and published masses: Ceto/Phorcys (0.64$^{+0.16}_{-0.13}$ g cm$^{-3}$), Typhon/Echidna (0.36$^{+0.08}_{-0.07}$ g cm$^{-3}$), and Eris/Dysnomia (2.40$^{+0.46}_{-0.37}$ g cm$^{-3}$).}
      
      \item{Beaming factor ($\eta$), determined from thermal fits of 7 objects, range from 1.02 to 1.48, with a weighted mean obtained of 1.14 $\pm$ 0.15, 
consistent with the mean value derived from \textit{Spitzer}, 1.20 $\pm$ 0.35 (Stansberry et al. 2008). These low $\eta$ values are compatible with low 
thermal inertias, presumably due to high porosity of regolith-like  surfaces.}
      
      \item{We obtain for the first time the albedo and size of the very large SDO 2007 OR$_{10}$: $D=$ 1280 $\pm$ 210 km, $p_R= 18.5^{+7.6}_{-5.2}$ \%. Along with
its surface composition, this indicates that the object is Quaoar-like.}
      
      \item{Our size/albedo estimations for the Eris/Dysnomia system ($D= 2454\pm117$ km, $p_V= 84.5 \pm 8.8$ \%) are compatible within error bars
with occultation results by Sicardy et al. (2011), but further analysis is needed.}

      \item{We find a significant correlation between albedo and diameter (more reflective objects being bigger), probably 
due to the fact that large objects can retain bright ices  more easily than small objects. Stansberry et al. (2008) obtained a similar result for a larger sample of TNOs, but they noted that the correlation is sensitive to changes in the classification used (i.e. MPC \textit{vs} DES; \cite{2005AJ....129.1117E})}.
      
	  \item{We also find positive correlations between albedo and perihelion distance, and between diameter and perihelion distance (brighter and bigger objects having
larger perihelia). The first correlation has been seen before based on {\em Spitzer} measurements (Stansberry et al. 2008) and interpreted in terms
of increased ice sublimation and/or increased space weathering at low heliocentric distances. 
It seems more difficult to find a physical explanation for the second correlation.}

  \end{itemize}


\begin{acknowledgements}

\textit{Herschel} is an ESA space observatory with science instruments provided by European led Principal Investigator consortia and with important participation from NASA. \textit{Herschel} data presented in this work were processed using “HIPE”, a joint development by the \textit{Herschel Science Ground Segment Consortium}, consisting of ESA, the NASA \textit{Herschel Science Center}, and the HIFI, PACS and SPIRE consortia. P. Santos-Sanz would like to acknowledge finantial support by the Centre National de la Recherche Scientifique (CNRS), also acknowledge Brett Gladman for his help with the dynamical classification of TNOs, and Daniel Hestroffer and Will Grundy for his help with the bulk density estimation of binary systems. M. Mommert acknowledges support through the DFG Special Priority Program 1385. Part of this work was supported by the German
\emph{Deutsches Zentrum f\"ur Luft- und Raumfahrt, DLR} project
numbers 50\,OR\,1108. We are grateful to Danielle Tibi for her unvaluable expertise on statistical aspects (Sect. B.1). J.L. Ortiz acknowledges the spanish grants AYA2008-06202-C03-01, AYA2011-30106-C02-01 and 2007-FQM2998. R. Duffard acknowledges financial support from the
MICINN (contract Ramon y Cajal).

\end{acknowledgements}


\Online

\begin{appendix} 

\section{Data reduction}

In this appendix, we describe the reduction process applied to the PACS data in order to obtain what we call ``single" and ``combined" maps.

\subsection{General PACS photometer data reduction}

As explained in the main text (see Sect. \ref{DataRed}) the data reduction of the maps was performed within HIPE (version 6.0.2044). We modified and adapted the standard HIPE scripts for the needs of our programme. Thus, the PACS photometer data is reduced from Level-0 to Level-2 using a modified version of the standard mini-scanmap pipeline. A major difference of our reduction chain compared with the standard pipeline is the application of frame selection on scan speed rather than on building block identifier (also known as BBID, which is a number that identifies the building block --a consistent part-- of an observation). It has been widely tested that the application of the scan speed selection increases the number of useable frames, and finally increases the SNR of the final maps by 10-30\% (depending on the PACS observation requests (AOR) setup and on the band used), which is especially important for our faint targets. 

Since our targets move slowly (a few $\arcsec/hour$), and the total observation time per scanning direction in one visit is between 10-25 min we do not correct for the proper motion. Performing this kind of correction would smear the background, and would make it impossible to make a correct background subtraction in the later stages (see subsection \ref{skysubtraction}). 

We use a two-stage high-pass filtering procedure, as in the standard mini scan-map pipeline. In the first stage a ``naive" map is created. This map is used in the second stage to mask the neighborhood of the source and the high flux pixels. In the second stage the vicinity of the
source has been masked (2 times the full width at half maximum -FWHM- of the beam of the actual band), as well as all pixels with fluxes above 2.5$\times$ the standard deviation of the map flux values. We used the standard high-pass filter width parameters of 15, 20 and 35. The final maps are created by the 
\textit{PhotProject()} task, using the default dropsize parameter. 

These first steps lead to the production of one ``single" map per visit, filter, and scan direction (i.e. in total 8 maps per object in the red, and 4 maps in the blue/green created using the \textit{MosaicTask()} in HIPE). The sampling of the single maps generated with these scripts are 1.1$\arcsec/pixel$, 1.4$\arcsec/pixel$, and 2.1$\arcsec/pixel$ for the blue (70 $\mu m$), green (100 $\mu m$) and red (160 $\mu m$) bands respectively. 

\subsection{Building final maps for photometry}
\label{skysubtraction}

\begin{figure}
   \centering
   
   \includegraphics[height=5.5cm]{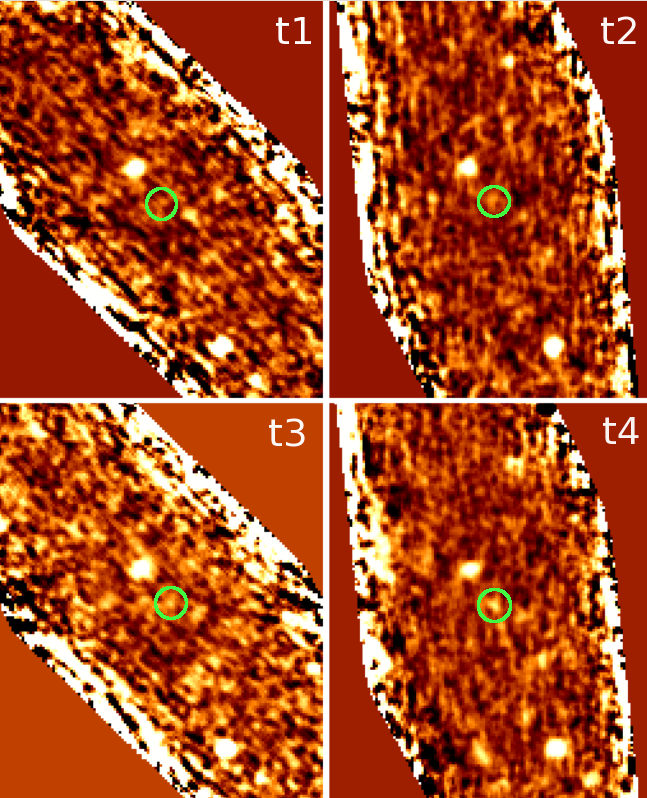}
   \includegraphics[height=5.5cm]{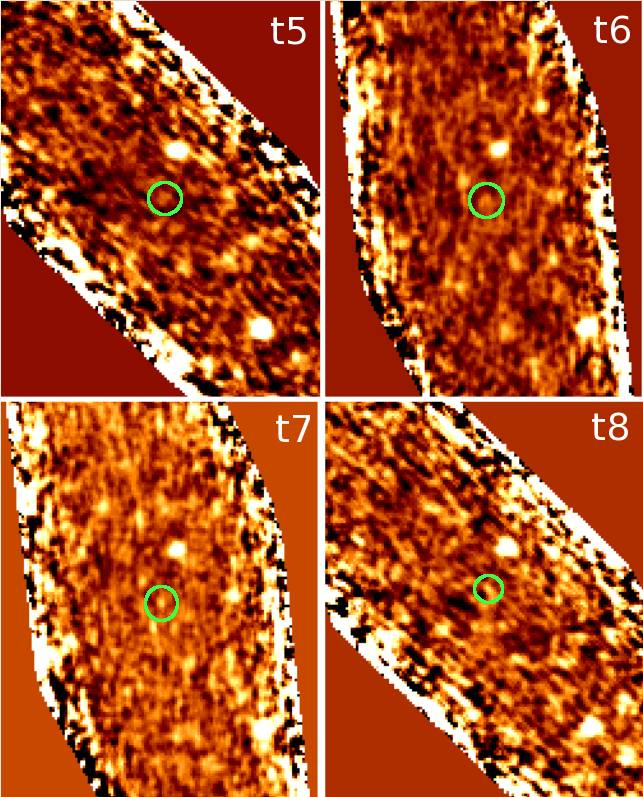}
   
      \caption{\textbf{Left four panels:} Typhon red band 1st visit single maps (t1 to t4). \textbf{Right four panels:} Typhon 2nd visit single maps (t5 to t8). t1, t3, t5, and t8 were observed with a scan angle of $110\degr$, and  t2, t4, t6, and t7 with a scan angle of $70\degr$ with respect to the detector array. The green circle marks the object position.}
         \label{typhon1}
   \end{figure}

We then use these single maps to generate the final ``combined" maps on which the photometry will be performed. Essentially, to generate the combined 
maps, we determine the background map, subtract it from each single map, and finally co-add all the background-subtracted maps. A detailed description of this process is as follows:

i) We have initially a set of 8 (red) or 4 (blue and green) single maps taken on different dates. Considering the red maps, let us call
t1 to t4 (t5 to t8) the dates corresponding to the first (second) visit. t1-t4 and t5-t8 are separated by typically 1-2 days so that the target motion (typically 30-$50\arcsec$ as described in Sect. \ref{observations} of the main text) produces a significant change in
the observed sky field (see Figure  \ref{typhon1}).

\begin{figure}
   \centering

   \includegraphics[height=5.5cm]{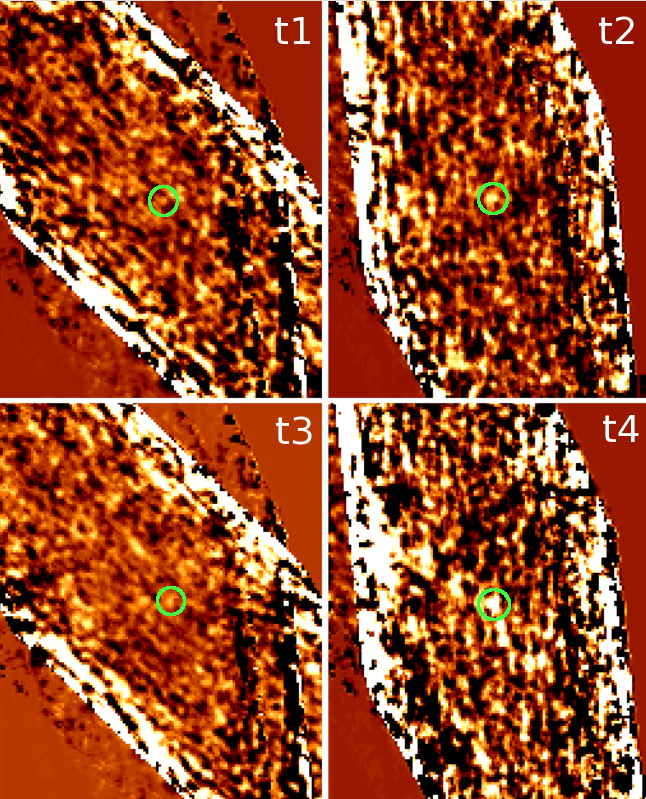}
   \includegraphics[height=5.5cm]{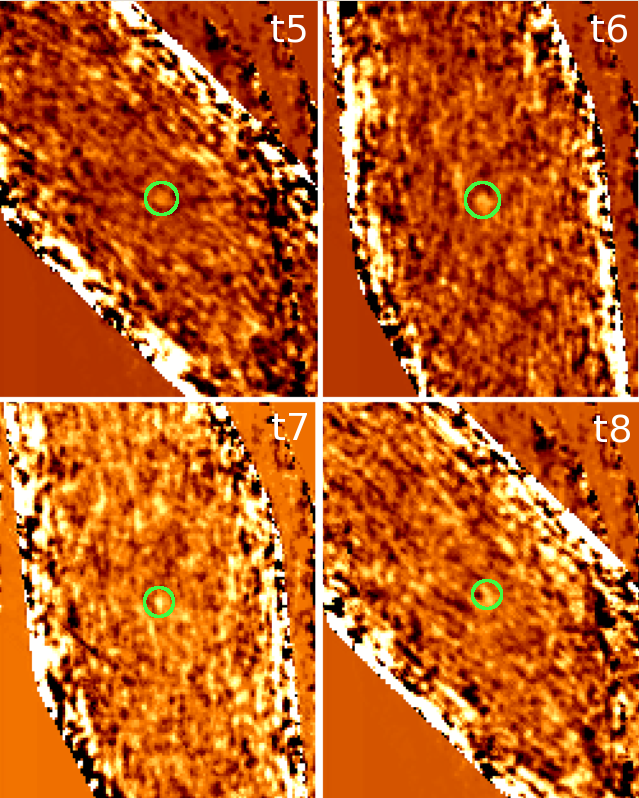}

      \caption{\textbf{Left:} Same as Figure 1 for the  background-subtracted maps. }

         \label{typhon2}
   \end{figure}

\begin{figure}
   \centering
   \includegraphics[width=9cm]{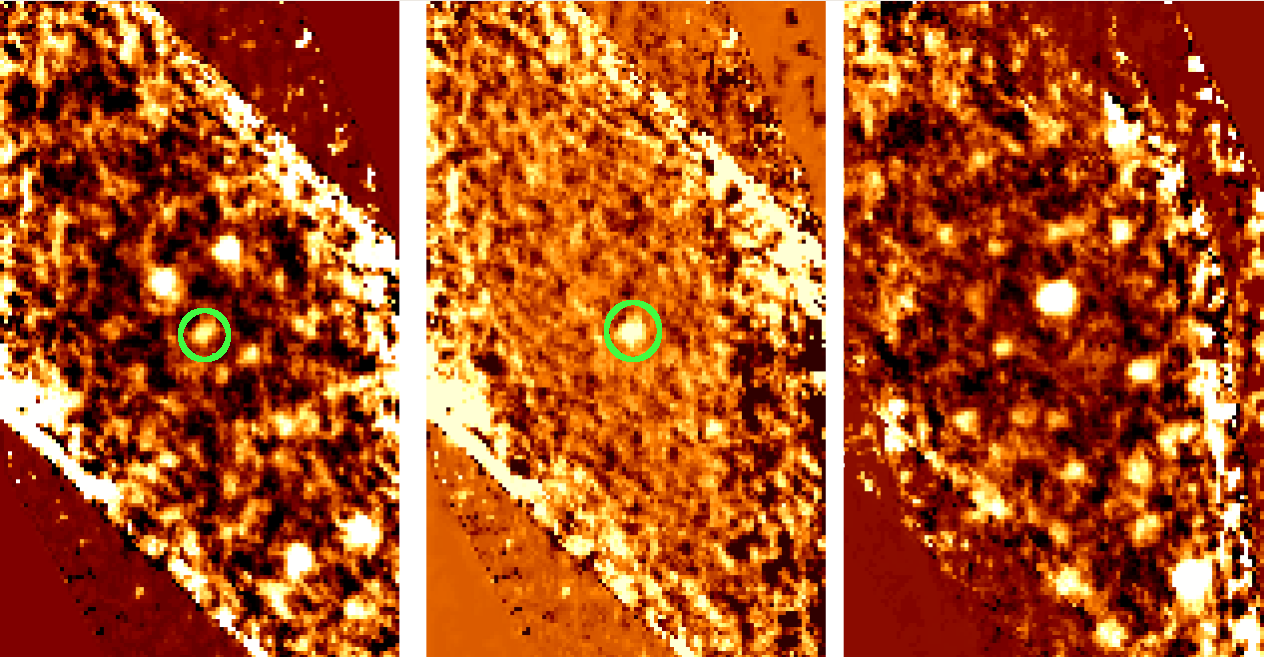}
      \caption{\textbf{Left:} Typhon simple coadded map in red band. Note that in this case the background is the average (centered in the target) of backgrounds in maps t1-t8 of Figure \ref{typhon1}, so that the bright background sources appear twice. Green circle marks the object position. \textbf{Center:} Typhon background-subtracted coadded map in red band. \textbf{Right:} Background map (target masked).}
         \label{typhon3}
   \end{figure}

ii) A background map is generated using the single maps. To do this we ``mask'' the target in each of the 8 (4) single maps and co-add the maps in the sky coordinate system. This step produces a background map with high SNR without the target (see the right panel in Figure \ref{typhon3} for the Typhon case at 160 micron).

iii) The background map is subtracted from the single maps (Figure  \ref{typhon1}), producing 8 (4) single maps with background removed. We call these images ``background-subtracted maps'' (Figure \ref{typhon2}).

iv) Finally the background-subtracted maps are co-added in the target frame (center panel of Figure  \ref{typhon3}), producing the final combined map on which
photometry is performed.

Note that in step ii) before combining the images of the two epochs, we need to calculate the optimal matching offset between the coordinate systems of the two maps since, when we use the original World Coordinate System (WCS) information stored in the file headers, artifacts of bright sources remain in the background subtracted image. The best offset position was found in the following way: we added a small offset --in steps of 0.5$\arcsec$ in both the R.A. and DEC directions-- to the appropriate original WCS parameters in the file headers, and performed the background subtraction. For each of these maps the standard deviation was calculated for pixels with a coverage value of at least 35 \% of the maximum coverage of the map. The offset values related to the lowest standard deviation correspond to the best matching of the two backgrounds. After this procedure the artifacts of bright sources (e.g. pairs of positive-negative spots) completely disappeared (or, at least, were below detection), and the noise in the background decreased notably. This method was performed on the 160\,$\mu$m band measurements, where the effect of the background is the largest. However, it has been checked in a series of measurements that the same offsets can be applied to the 70 and 100\,$\mu$m images as well.

An alternate, simpler method, to obtain final maps is to co-add directly the original single maps (i.e. not background-corrected) in the target frame. We call this the simple coaddition method (left panel of Figure \ref{typhon3}). This method is obviously less optimal than the previous one in terms of SNR, but provides a useful test to demonstrate that the background subtraction is not introducing any spurious effects. 

A more detailed description about the whole PACS data reduction process will be published in Kiss et al. (\textit{in prep.}-a).

\begin{figure}
   \centering
    \includegraphics[width=8.5cm]{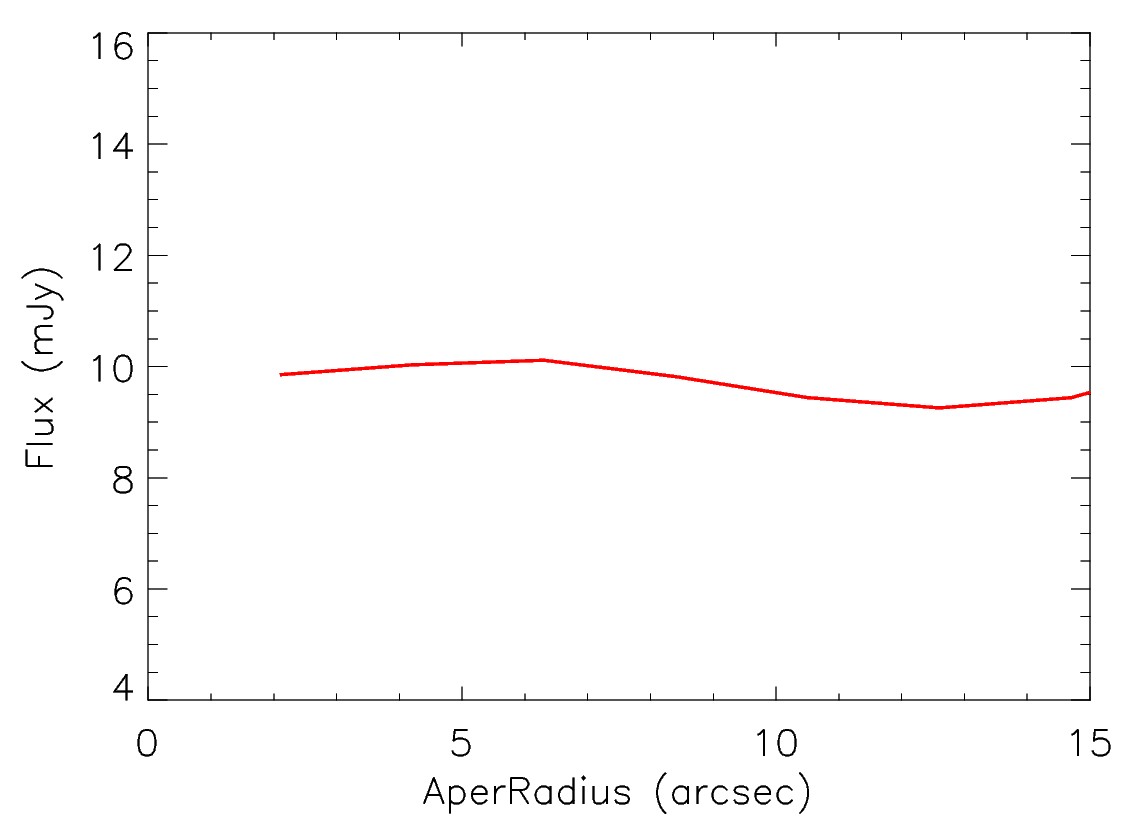}
      \caption{Aperture-corrected curve-of-growth obtained from Typhon background-subtracted coadded map in red band (Figure \ref{typhon3}-center).
              }
         \label{Typhon}
   \end{figure}

\end{appendix}


\begin{appendix} 

\section{Statistical aspects}

\subsection{``Rescaled error bar" approach}
\label{infladas}

As indicated in the main text (see Sect. \ref{MCapproach}), error bars on the fitted parameters 
(diameter, albedo and beaming factor) are determined through a 
Monte-Carlo approach which essentially consists of randomly building synthetic datasets using uncertainties in the measured fluxes.
However, when the measurement errors bars are too small, or similarly when the model
is not completely adequate (both cases being indicated by a ``poor model fit"), the described Monte-Carlo approach will
underestimate the uncertainties on the solution parameters. In this case, we adopted a 
``rescaled error bar" approach described hereafter.

Quantitatively, the fit quality is determined by the value of 
the reduced $\chi^2$:

\begin{equation}
	\chi_\mathrm{red}^2 = \frac{\chi^2}{\nu} = \frac{1}{\nu} \sum {\frac{(O - M)^2}{\sigma^2}}
\end{equation}

where $O$, $M$ and $\sigma$ are the observed, modelled and error bar flux values, and $\nu$
is the degree of freedom ($\nu$ = $N$-1 for fixed-$\eta$, and $\nu$ = $N$-2 for floating-$\eta$, 
where $N$ is the number of thermal wavelengths available).  While $\chi_\mathrm{red}^2 \sim$ 1 indicates
a good fit, $\chi_\mathrm{red}^2 >>$ 1 indicates a poor fit. The idea is therefore to
empirically ``rescale" (uniformly, in the lack of a better choice) all error bars $\sigma$ by 
$\sqrt{\chi_\mathrm{red}^2}$ before running the Monte-Carlo approach. This method leads to much more 
realistic error bars on the solution parameters. However, the range of $\chi_\mathrm{red}^{2}$  for which this 
approach is warranted depends on the number of data points, as for few observations 
$\chi_\mathrm{red}^2$ does not have to be closely equal to 1 to indicate a good fit.
Specifically, the variance of the variance estimator of a distribution of $N$ points picked from a Gaussian distribution with 
dispersion = 1 is itself a Gaussian with a mean equal to unity and a standard deviation equal to $\sqrt{2/N}$.
If we had for example $N$ = 1000 observations, 68.2 \% of the good fits have a reduced   
$\chi^2$ between 0.955 and 1.045, i.e. $\chi_\mathrm{red}^2$ is strongly constrained to 1; 
for $N$ = 5 (resp. 3) observations in contrast, this range is
much broader, 0.368-1.632 (resp. 0.184-1.816). This means for example that for 3 data
points (PACS only), a reduced $\chi^2$ of e.g. 1.6 is acceptable.  We considered these $N$-dependent 
limits as the thresholds beyond which the fits are statistically bad and the error bars need to be rescaled. 
Note finally that the rescaling approach is merely an ``operative" method to avoid unrealistically
low error bars on the fitted diameter, albedo and beaming factor, and that error bars shown in Fig. \ref{PACS_fits} and \ref{MIPS_PACS_fits} 
of the main text are the original, not rescaled measurement uncertainties. 

\subsection{Spearman-rank correlation with error bars}
\label{spearmanconerrors}

As described in the main text the Spearman-$\rho$ correlation test is distribution-free 
and less sensitive to outliers than most other methods but, like others, treats data-points as `exact' and
does not take into account their possible error bars. Any variations in
the measured data points within their error bars, may change the
correlation coefficient, however. Furthermore, each correlation
coefficient has its own confidence interval, which depends on the number
of data points and on the magnitude of the correlation. To account for these effects we used the following three procedures:

\begin{enumerate}

\item To estimate the most probable correlation coefficient given the
error bars, we generate 1000 samples of data points, building each
synthetic dataset from its associated probability function. When errors
bars are symmetric this probability function is considered to be Gaussian
and the error bar correspond to one standard deviation. When errors bars
are asymmetric, we fit a lognormal probability function such that the
observed value corresponds to the distribution's mode, and the interval
between the upper and lower limits of the observation corresponds to the
shortest interval containing $68.2\%$ of all possible values. The
resulting distribution of correlation values is not Gaussian, but its
Fisher-transform $z= \arg \tanh(\rho)$ is. So, we can determine the most
probable correlation value, $\langle \rho \rangle$ from our Monte-Carlo
simulations, and its upper and lower limits
(${+\sigma_{MC}}/{-\sigma_{MC}}$), which are not necessarily symmetric
after the reconversion using $\rho=\tanh(z)$ (see
\cite{2004Icar..170..153P} for more details). It is noticeable (and
expected) that observational error bars `degrade' the correlation. The
approximate significance p-value, or confidence level (CL), of $\rho$ can
be computed from $t=\rho \sqrt{(n-2)/(1-\rho^2)}$, which follows a
t-Student distribution with $n-2$ degrees of freedom. However, when $n\lesssim15$, using $t$ provides a slight overestimation of the significance. We have
compared the previous calculation of the significance with exact values
tabulated by Ramsey (1989) and computed adjustments required to obtain a
more accurate approximation of the $\rho$'s and p-values in the case of
low $n$.

\item The confidence level of a given $\rho$ results from testing the
hypothesis `the sample is correlated' against the hypothesis `the sample
correlation is zero'. Knowing the confidence interval (CI) within which
the correlation $\rho$ value of the parent population lies may be more
informative. For example, suppose we have $\rho=0.7$ with a $3\sigma$
confidence level; we would conclude that the sample is correlated. But if
the $68\%$ confidence interval of the correlation were say $[0.3, 0.9]$,
we would be unsure if the correlation was very strong or rather weak.
To estimate the shortest ${(+\sigma_B}/{-\sigma_B)}$ interval containing
$68.2\%$ of the population's $\rho$ ({\it i.e.} the equivalent to the
Gaussian $1\sigma$ interval), we used 1000 bootstrap extractions from the
data-sample computing this range as we did for the previous
item (e.g. \cite{Efron1993}).

\item Finally, to be fully correct, one should perform the 1000 bootstraps
on each one of our 1000 Monte-Carlo simulations to obtain the true
$68\%$ confidence interval of $\langle \rho \rangle$, a computationnally
heavy task. Fortunately, the combination of the two effects can be
obtained by quadratically adding
the standard deviations of the Gaussian distributions of the
Fisher-transform of the Monte-Carlo simulations and the bootstraps, i.e.
$\sqrt{\sigma_{MC}2+\sigma_B2}$, which, after reconversion to $\rho$,
will give our final best estimate for the $68\%$ confidence interval noted
as $\langle \rho \rangle^{+\sigma}_{-\sigma}$.

\end{enumerate}

\end{appendix}

\end{document}